\documentclass[fleqn,usenatbib,useAMS]{mnras}

\usepackage{graphicx}	
\usepackage{amsmath}	
\usepackage{amssymb}	
\usepackage{multicol}        
\usepackage{bm}		
\usepackage{pdflscape}	

\usepackage[T1]{fontenc}
\usepackage{ae,aecompl}

\usepackage{newtxtext,newtxmath}

\title[$^7$Be II in the SMC]{Detection of $^7$Be II in the Small Magellanic Cloud}

\author[L. Izzo et al.]{Luca Izzo$^{1}$\thanks{Contact e-mail: luca.izzo@nbi.ku.dk, paolo.molaro@inaf.it}, Paolo Molaro$^{2,3}$, Gabriele Cescutti $^{2,3,4}$, 
\newauthor
Elias Aydi$^{5}$, Pierluigi Selvelli$^{2}$, Eamonn Harvey$^{6}$, Adriano Agnello$^{1}$,
\newauthor
Piercarlo Bonifacio$^{7}$, Massimo Della Valle$^{8,9}$, Ernesto Guido$^{10}$, Margarita Hernanz$^{11}$
\\
$^{1}$DARK, Niels Bohr Institute, University of Copenhagen, Jagtvej 128, 2200 Copenhagen\\
$^{2}$INAF, Osservatorio Astronomico di Trieste, Via Tiepolo 11,  I-34143 Trieste, Italy\\
$^{3}$IFPU, Istitute for the Fundamental Physics of the Universe, Via Beirut, 2, I-34151 Grignano, Trieste, Italy\\
$^{4}$INFN, Sezione di Trieste, Via A. Valerio 2, I-34127 Trieste, Italy\\
$^{5}$Center for Data Intensive and Time Domain Astronomy, Department of Physics and Astronomy, Michigan State University, East Lansing, MI 48824, USA\\
$^{6}$Astrophysics Research Institute, Liverpool John Moores University, IC2 Liverpool Science Park, Liverpool, L3 5RF, UK\\
$^{7}$ GEPI, Observatoire de Paris, Universit{\'e} PSL, CNRS, Place Jules Janssen, 92195 Meudon, France\\
$^{8}$ Capodimonte Astronomical Observatory, INAF-Napoli, Salita Moiariello 16, 80131-Napoli, Italy\\
$^{9}$ ICRANet, Piazza della Repubblica 10, I-65122 Pescara, Italy\\
$^{10}$Telescope Live, Spaceflux Ltd, 71-75 Shelton Street, Covent Garden, London, WC2H 9JQ, UK\\
$^{11}$ Institute of Space Sciences (ICE, CSIC) and IEEC, Campus UAB, Cam{\'i} de Can Magrans s/n, 08193 Cerdanyola del Valles (Barcelona), Spain\\}

\date{}

\pubyear{2019}
\hypersetup{draft}
\begin{document}
\label{firstpage}
\pagerange{\pageref{firstpage}--\pageref{lastpage}}
\maketitle

\begin{abstract}
We  analyse  high resolution spectra of two classical novae that exploded in the Small Magellanic Cloud.  $^7$\ion{Be}{ii} resonance transitions are detected in both ASASSN-19qv and ASASSN-20ni novae. This is the first detection outside the Galaxy and confirms that thermo-nuclear runaway reactions, leading to the $^7$Be formation, are effective  also in the low metallicity regime, characteristic of the SMC. Derived yields are of  N($^7$Be=$^7$Li)/N(H) = (5.3 $\pm$ 0.2) $\times$ 10$^{-6}$ which are a factor 4 lower than the typical  values of the Galaxy. 
Inspection of two historical novae in the Large Magellanic Cloud  observed with IUE in 1991 and 1992 showed also the possible presence of $^7$Be and similar yields. For an ejecta of 
 $M_{H,ej} =$ 10$^{-5}$ M$_{\odot}$, the amount of $^7$Li produced  is of    $M_{^7 Li} = (3.7 \pm 0.6) \times 10^{-10}$ M$_{\odot}$ per nova event. Detailed chemical evolutionary model for the SMC shows that novae could have made an amount  of lithium in the SMC corresponding to a fractional abundance of A(Li) $\approx$  2.6. Therefore, it is argued that a comparison with the abundance of Li in the SMC, as measured by its interstellar medium, could
effectively constrain the amount of the initial abundance of primordial Li, which is currently controversial.
\end{abstract}
\begin{keywords}
{stars: individual: ASSASN-19qv, ASASSN-20ni; stars: novae
-- nucleosynthesis, abundances; Galaxy: evolution -- abundances}
\end{keywords}



\section{Introduction}

Lithium is the only {\it metal} element produced during the Big-Bang nucleosynthesis (BBN) due to the lack of stable nuclei with mass number eight \citep{Fields2014}. The element abundances predicted by the standard BBN theory for the baryonic density coming from the \emph{Planck} mission agree well with those observed, except for  $^7$Li  \citep{Fields2011,Coc2014}. Indeed, the abundance of lithium measured in  the low-metallicity Galactic halo stars is A($^7$Li) = log[N$(^7Li)/N(H)] +12 = 2.25$ \citep{Spite1982,Sbordone2010,Bonifacio2015},  which is  $\sim$ 3 times below the estimate of the standard cosmological model  $A(^7Li) = 2.72 \pm 0.06$ \citep{Cyburt2016}. The latter value depends on the baryon-to-photons ratio $\eta = \frac{N_b}{N_{\gamma}} \propto \Omega_B h^2$, with $\Omega_b$ the cosmological baryon density and $h$ the dimensionless hubble parameter \citep{Planck2016}. This problem is also known as the {\it Cosmological Lithium problem} \citep{Fields2014}. A possible solution can be ascribed to convective diffusion in the pre-main sequence phase as well as during the lifetime of these halo stars \citep{Fu2015} or to new physics beyond the standard model. On the other hand,  the  young stellar populations in our Galaxy  show  Li-abundances  four times greater than the SBBN estimate and more than  one order of magnitude  greater than the halo stars \citep{Spite1990,Lambert2004,Lodders2009,Ramirez2012,Fu2018}. 
The evidence of a growth requires the existence of additional lithium factories. In the last decades several astrophysical Li sources have been  proposed, like Galactic cosmic-rays, AGB stars, low-mass Carbon stars, type II supernovae and classical novae \citep{DAntonaMatteucci1991,Romano1999,Prantzos2012, Matteucci2021a}. The recent detection in the outburst  spectra of  classical novae of $^7$Li and $^7$\ion{Be}{ii}, an isotope whose unique decay channel is into lithium through electron capture,  have confirmed these objects as Li producers. The corresponding yields inferred  have placed  nova  explosions as the main lithium factories in the Galaxy. The time scales involved also match,  as shown by detailed Galactic chemical evolution \citep{Izzo2015,Tajitsu2015,Molaro2016,Izzo2018,Molaro2020,Cescutti2019,Grisoni2019,Matteucci2021a}.

Classical Novae (CNe) are stellar explosions originating from  a  white dwarf that accretes matter from a late-type main sequence, or in some cases a red giant  companion \citep{BodeEvansbook}. The matter accreted on to the white dwarf surface piles up leading to an increase of  pressure and temperature  until CNO thermo-nuclear reactions ignite \citep{Gallagher1978} which leads to an explosive ejection of the accreted layers into the interstellar medium \citep{Gehrz1998}. During this thermo-nuclear runaway process (TNR, \citealp{Starrfield1978}), we witness the formation of the $^7$Be isotope through the $^3$He($\alpha$,$\gamma$)$^7$Be process. The synthesized beryllium decays into lithium through electron-capture with a half-life time decay of $\sim$ 53 days \citep{CameronFowler1971}. Given that $^7$Li is very easily destroyed in almost every astrophysical process, $^7$Be has to be transported to zones that are cooler than those where it was formed, with a timescale shorter than its decay time, in order to be detected. This beryllium transport mechanism, as first suggested by \citet{Cameron1955}, requires a dynamic situation that is encountered so far only in asymptotic giant branch (AGB) stars and novae.

With an absolute magnitude at maximum that ranges  between $V = -10$ mag and $V = -6$ mag \citep{DellaValle2020}, CNe can be  observed also in nearby galaxies, in particular in the nearby Magellanic clouds. These two Milky Way galaxy satellites  are characterised by a low metallicity ($\sim$0.5 $Z_{\odot}$ for the LMC and $\sim$0.2 $Z_{\odot}$ for the SMC, \citealp{Madden2013}). 
Only in recent years, thanks to  high-resolution spectrographs mounted at large telescopes, it was possible  to detect the very weak interstellar line of $^7$Li $\lambda$ 670.8nm towards a star belonging to the SMC \citep{Howk2012}. The measurement of its abundance, $A(^7Li) = 2.68 \pm 0.16$ taken at face value is very close  with the predictions from the standard BBN, $A(^7Li) = 2.72 \pm 0.06$ \citep{Cyburt2016}.


Following the  detection  of $^7$Be in the ejecta of several classical Galactic novae \citep{Tajitsu2015,Molaro2016,Izzo2018,Molaro2020}, we report here the attempts to observe this isotope in  extragalactic classical novae. After the 2016 outburst of the SMC Nova 2016-10a, also known as MASTER OT J010603.18-744715.8 \citep{Aydi2018}, we had to wait until July 4, 2019 to observe another bright nova in the SMC, ASASSN-19qv (SMCN-2019-07a), and  one more year  to observe ASASSN-20ni (AT2020yeq). In this work we present the first extragalactic $^7$\ion{Be}{ii} detection  in high-resolution spectral observations of the novae ASASSN-19qv and ASASSN-20ni. The   implications for  the  chemical evolution of lithium in the SMC are then discussed.

\section{Observations}

\subsection{ASASSN-19qv}

\begin{figure}
 \includegraphics[width=\columnwidth]{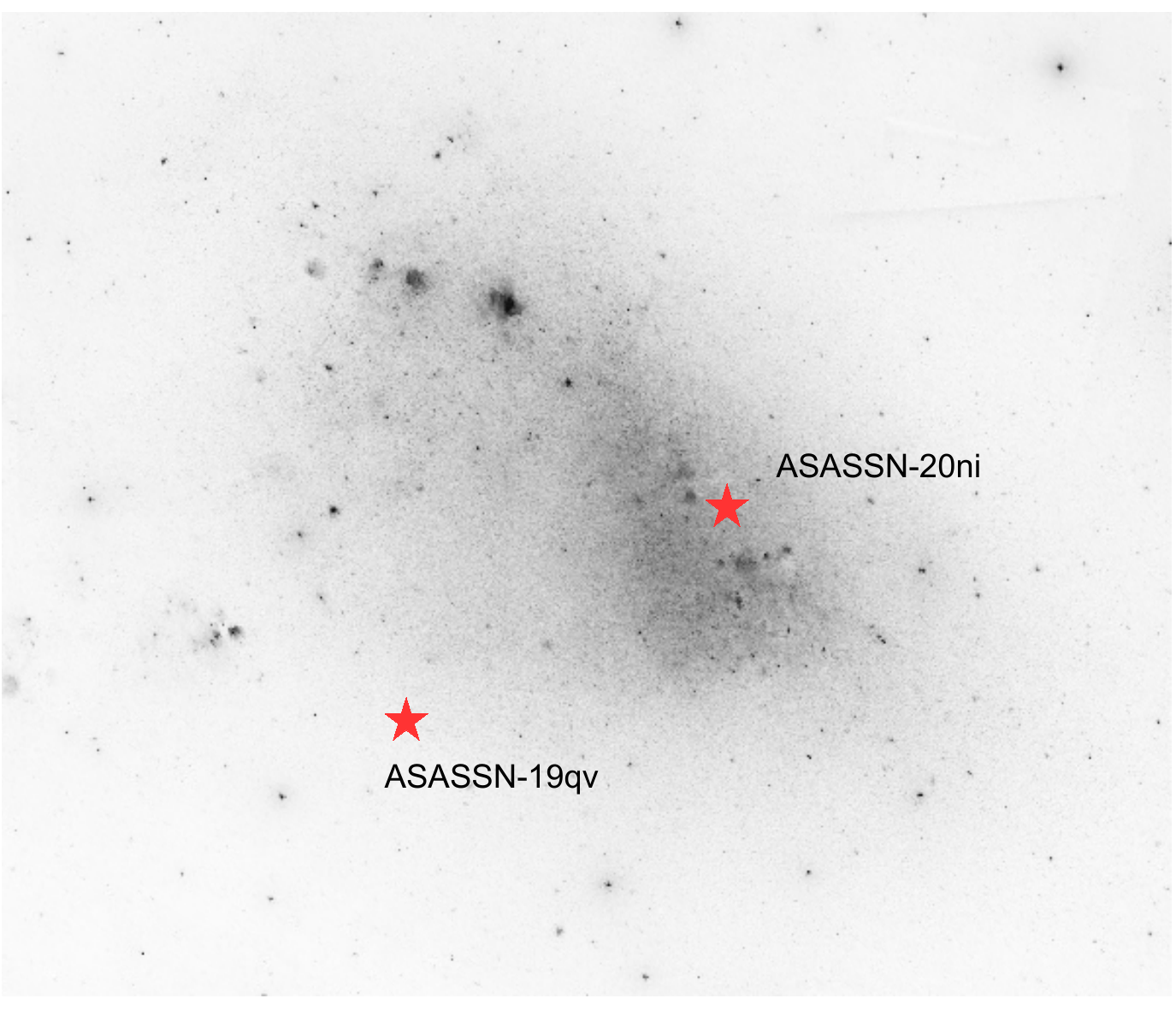}
 \caption{The DSS2 image of the SMC obtained by the Anglo-Australian Observatory (AAO) with the UK Schmidt Telescope. The positions of the two novae are marked with a red star. }
 \label{fig:8}
\end{figure}

\begin{figure*}
 \includegraphics[width=0.92\columnwidth]{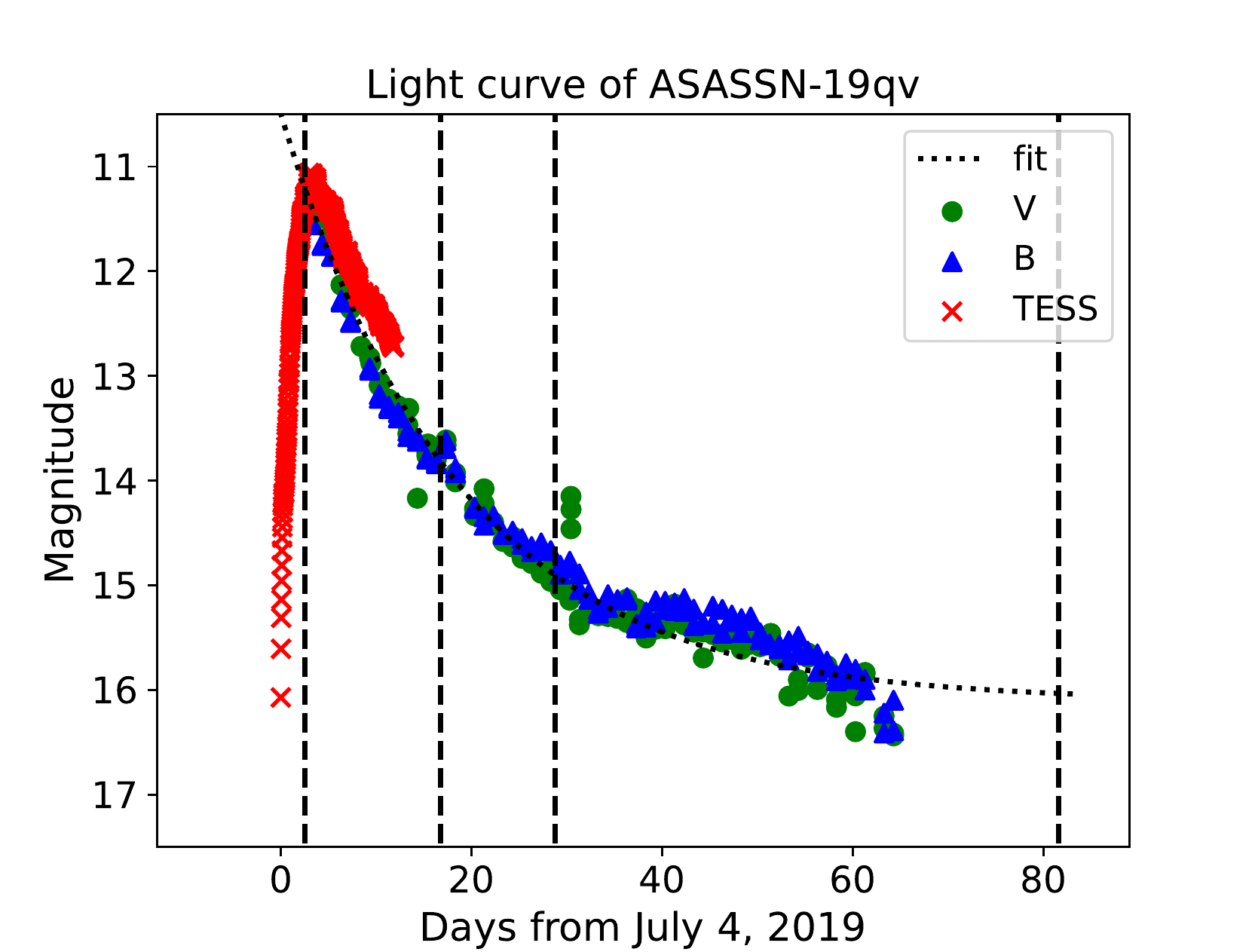}
 \includegraphics[width=0.99\columnwidth]{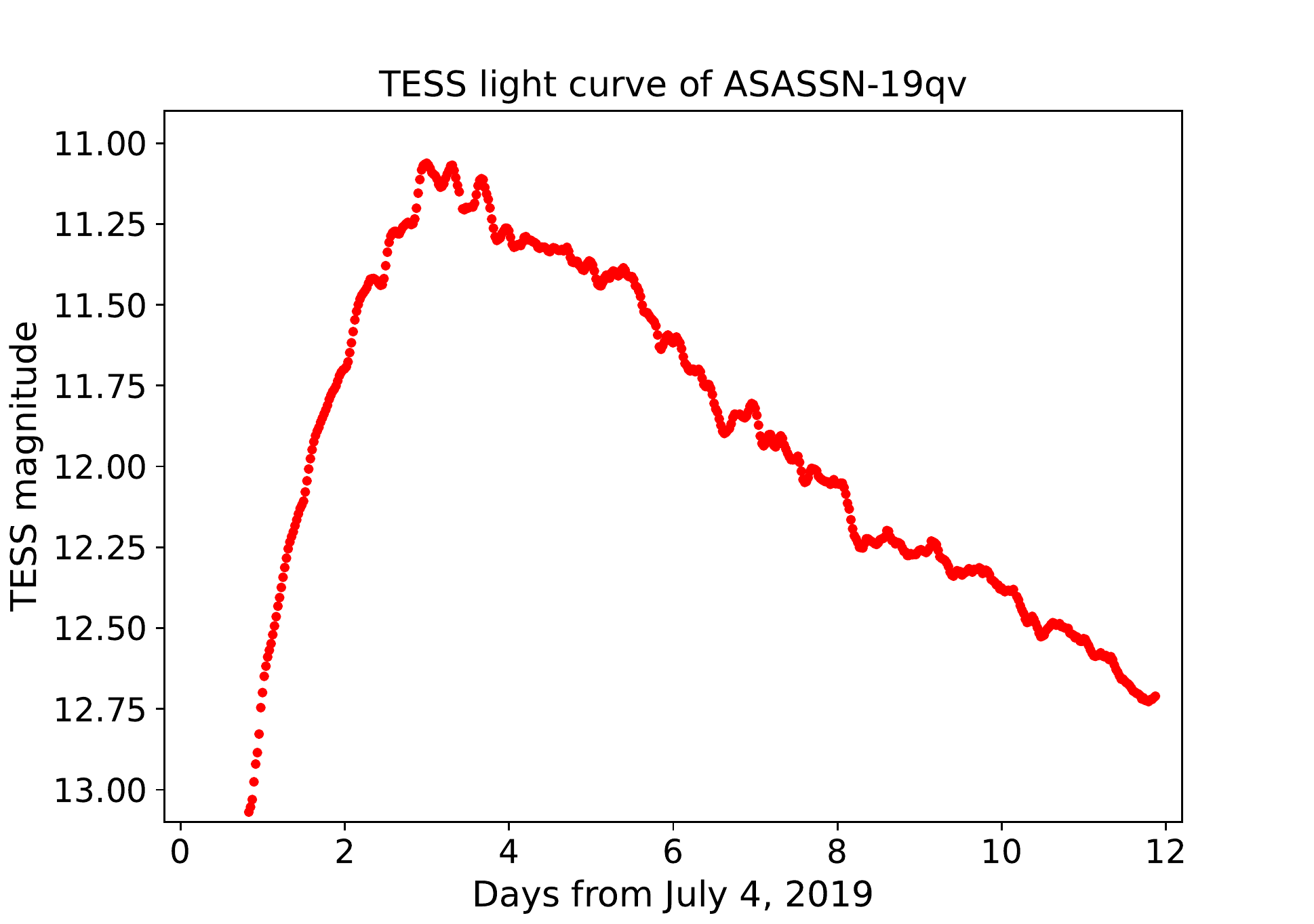}
 \caption{(Left panel) The light curve of ASASSN-19qv as obtained from the AAVSO data archive \citep{Kafka2021} in the $B$ and $V$ filters and using the early TESS data, as described in the text. Dashed lines correspond to the spectral epochs presented in this work. The dotted curve represents the best-fit that we have found using an exponential decay function. (Right panel) An in-depth view of TESS data. The light curve shows small fluctuations in magnitude with amplitude of 0.1-0.2 mag.}
 \label{fig:1}
\end{figure*}

\begin{figure}
 \includegraphics[width=\columnwidth]{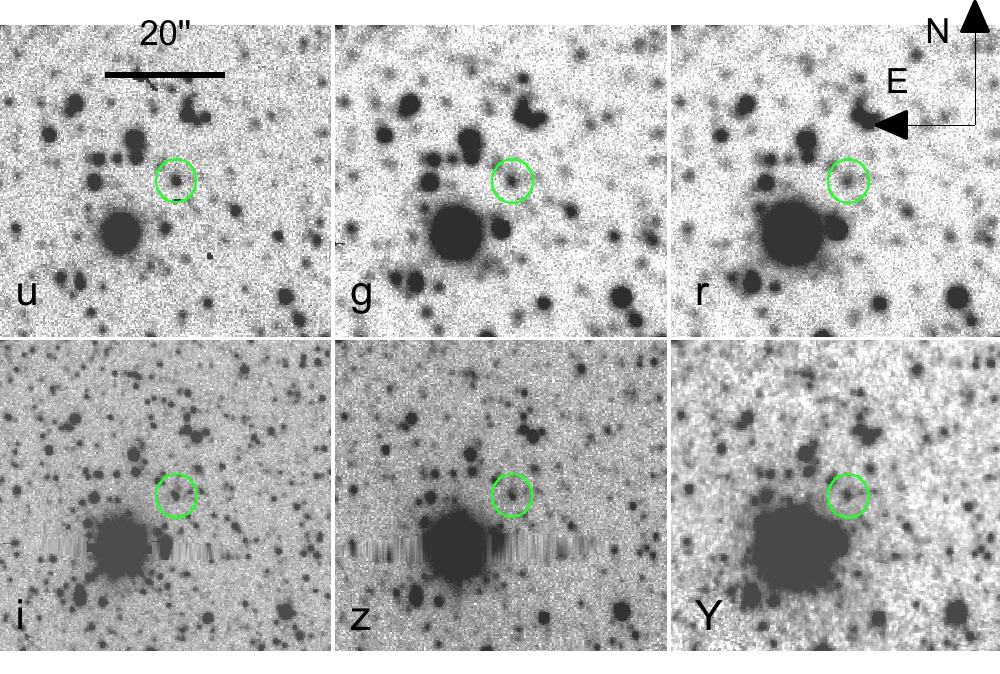}
 \caption{Multi-images of the field of view around ASASSN-19qv as observed in the $ugriz$-bands by the SMASH survey \citep{Nidever2017} and in the $Y$-band as observed by the VISTA Magellanic Cloud survey \citep{Cioni2011}. The position of the nova is marked with a green circle and it shows the counterpart reported in the text.}
 \label{fig:1b}
\end{figure}

The classical nova ASASSSN-19qv was discovered by the ASAS-SN survey \citep{Shappee2014} on  July 4, 2019 as a new transient of $g$ = 14.2 mag in the direction of the SMC as shown in Fig. \ref{fig:8}. There is a source in the Gaia DR2 (ID 4685624636344633728) at the position of the nova for which the reported parallax is negative, suggesting a very distant object, in agreement with being as distant as the SMC. The Gaia $G$ magnitude for this source is $G = 20.68$ mag. The field of view of ASASSN-19qv was observed by the SMASH survey \citep{Nidever2017} in the $ugriz$ filters. There is a source at the position of ASASSN-19qv in all the filter images, which is slightly extended toward the NE direction, suggesting that this source is actually composed by two stars. Using nearby stars taken from the USNO B1 catalog we measure a magnitude for this source of $i = 20.0 \pm 0.5$ mag. We have also found a catalog photographic magnitude of $B_{j} = 20.65$ in the Guide Star Catalog release 2.3 \citep{Bucciarelli2008} for the source reported at the position of the nova. Additionally, the field of view of ASASSN-19qv was covered by the VISTA Magellanic Cloud survey \citep{Cioni2011} on Aug 16, 2014. One source in the Y-band  is coincident with the position reported for ASASSN-19qv, see fig. \ref{fig:1b}, for which we determine $Y = 21.3$ mag. 
However, the double nature of the SMASH source suggests that it could be a foreground faint Galactic star or possibly red giant in the SMC. High-resolution imaging combined with spectroscopic observations of this source during the quiescence phase will definitely reveal the real nature of the progenitor.

A first spectrum obtained two days after the discovery of the nova confirmed the transient as a classical nova thanks to the identification of P-Cygni lines of \ion{Fe}{ii}, \ion{O}{i} and \ion{Na}{i} in addition to Balmer lines, with expanding blue-shifted velocities of $\sim$ -900, -1000 km/s \citep{Aydi2019}. Spectroscopic observations obtained in the following days (Days 5 and 8, \citealp{Bohlsen2019a,Bohlsen2019b}) still showed the presence of \ion{Fe}{ii} absorption lines and the absence of higher ionization transition like \ion{He}{i}. We started to observe ASASSN-19qv with the UVES spectrograph at ESO Very Large Telescope (Program ID: 2103.D-5044, PI Izzo) after 16 days from  discovery. The wavelength range covered by UVES starts from 310 nm to 950 nm at a spectral resolution of $R = 40,000$. The following two epochs, Day 29 and Day 81, were obtained with X-shooter at ESO/VLT, covering a wider spectral range from 300 nm to 2500 nm and with a resolution variable from $R = 6,700$ for the VIS and NIR arms to $R = 8,900$ for the UVB arm. The detailed log of the observations is shown in Table \ref{tab:1}. 

\subsection{ASASSN-20ni}

ASASSN-20ni was also discovered by the ASAS-SN survey on October 26, 2020  as a new transient of $g$ $\sim$ 14.1 mag and was confirmed the following day when it increased in brightness to $g$ $\sim$ 12.2 mag \citep{Way2020}. The ASAS-SN survey also monitored the position of the sky where ASASSN-20ni was located in the previous six years, and no previous outburst from the nova progenitor brighter than $g >$ 16.5 mag were reported.

\begin{figure}
 \includegraphics[width=\columnwidth]{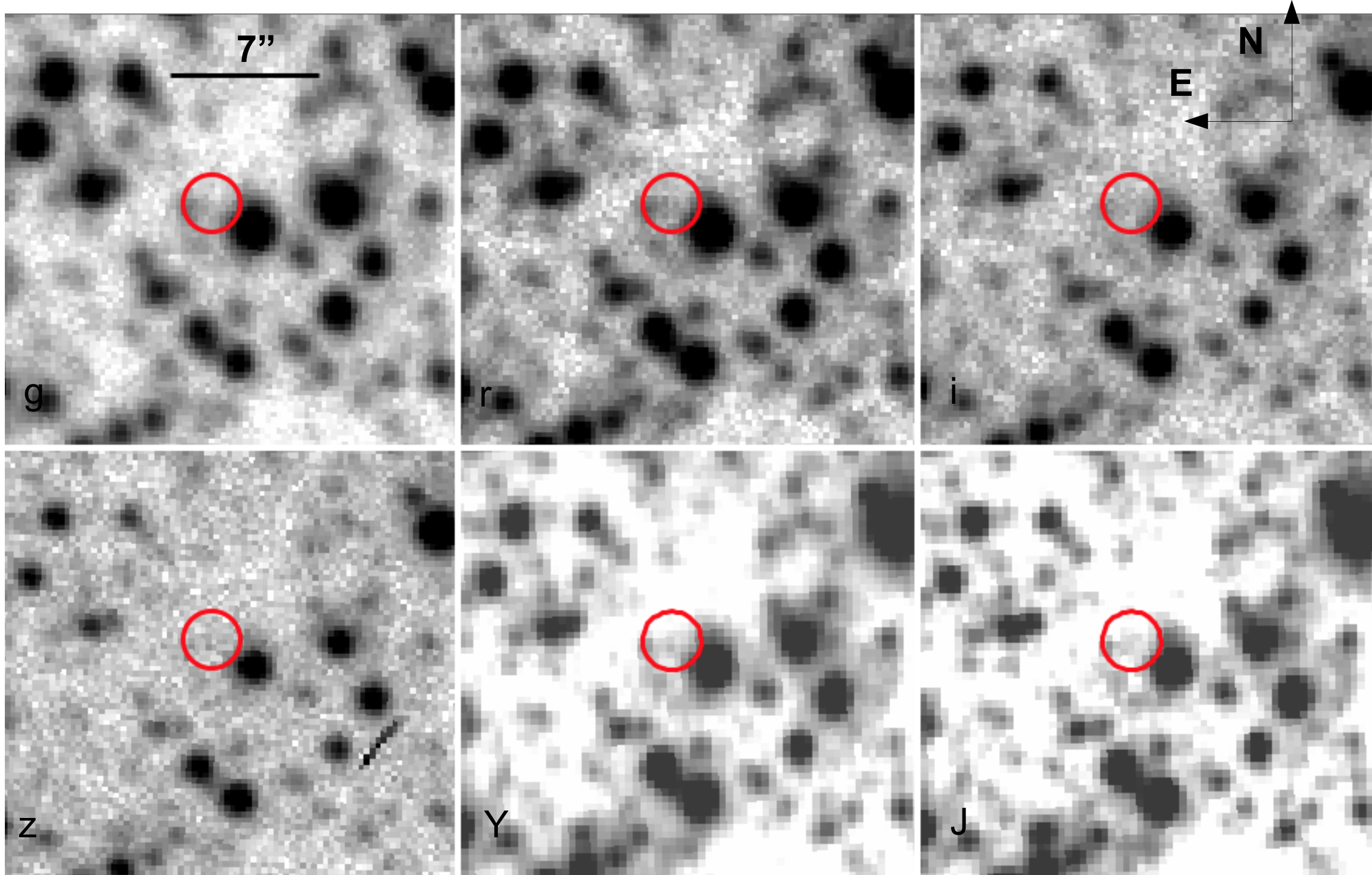}
 \caption{Multi-images of the field of view around ASASSN-20ni as observed in the $griz$-bands by the SMASH survey \citep{Nidever2017} and in the $YJ$-bands as observed by the VISTA Magellanic Cloud survey \citep{Cioni2011}. The position of the nova calibrated with the UVES acquisition image is marked with a red circle.}
 \label{fig:1c}
\end{figure}

We  searched for the possible presence of the nova progenitor in archival data surveys. To improve the astrometry from ASAS-SN we used the UVES acquisition image to calibrate astrometry at sub-arcsecond precision. No clear sources have been found in the SMASH survey and in the VISTA Magellanic Cloud survey down to a magnitude limit of Y $>$ 21.5 mag, despite the nova being 2 arcsec from a bright ($Y$ = 17.8 mag) star and 1.5 arcsec from another fainter source, see also Fig. \ref{fig:1c}.

The day following the discovery ASASSN-20ni was classified as a \ion{Fe}{ii} nova  using the Goodman spectrograph covering the wavelength range between 620 and 720 nm \citep{Aydi2020c}. The spectrum was characterised by P-Cygni emission line profiles for Balmer, Fe and N lines with absorption trough minimum at $v_{exp} \sim -500$ km/s. We started to observe ASASSN-20ni with the UVES spectrograph using a dedicated Discretional Director Time program (Program ID: 2106.D-5008(B), PI Izzo)  four days after the nova discovery.  We used different exposure times for the different UVES arms and dichroic configurations, in order to optimize the signal-to-noise in the near-UV range and at the same time avoid possible saturation from bright lines between 480 and 850 nm. The log of the observations is shown in Table \ref{tab:1}.

\begin{figure}
 \includegraphics[width=0.95\columnwidth]{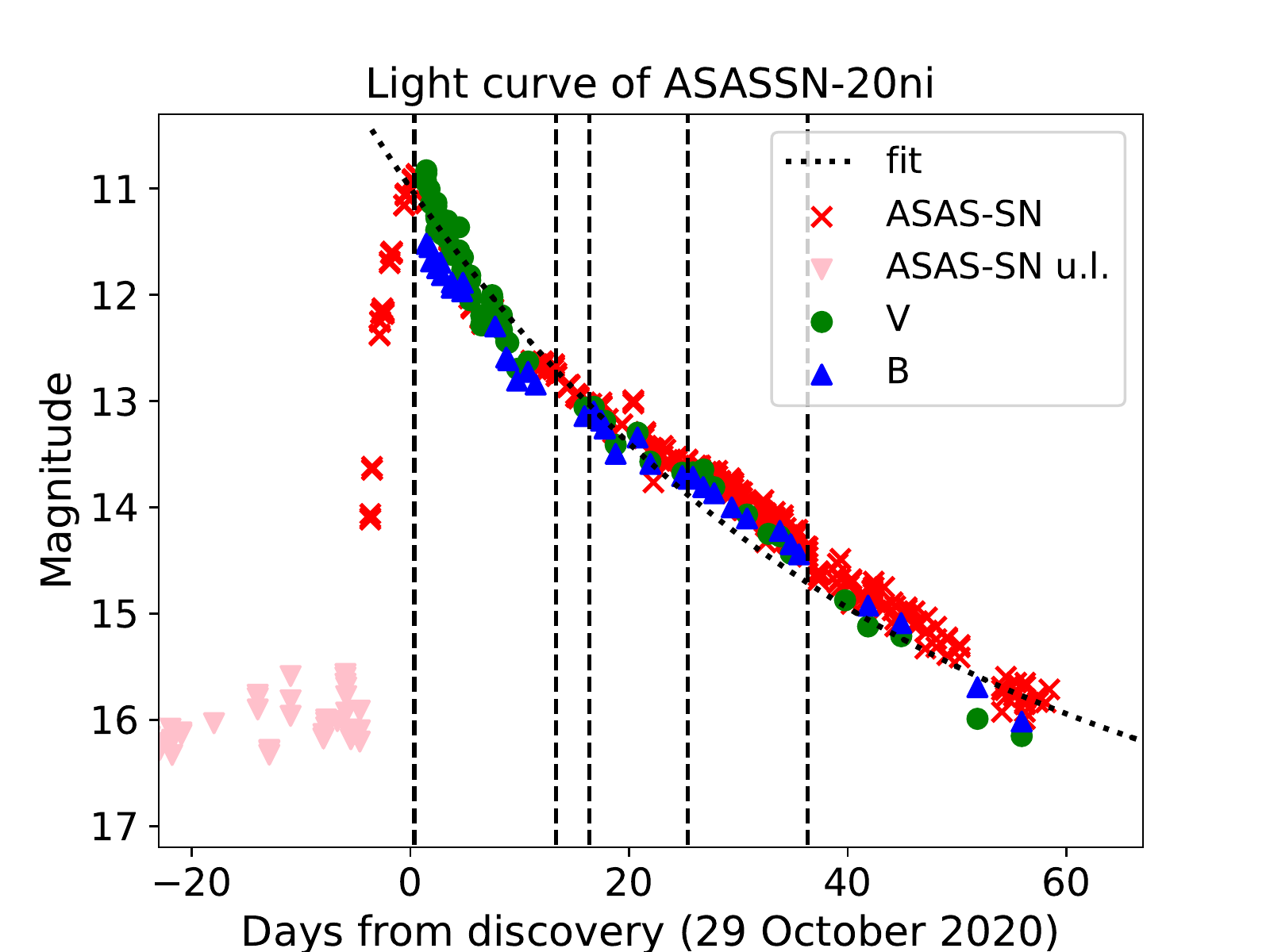}
 \caption{The light curve of ASASSN-20ni as obtained from the AAVSO data archive \citep{Kafka2021} in the $B$ and $V$ filters and using the ASAS-sn $g$ filter data, as described in the text. Dashed lines correspond to the spectral epochs presented in this work. The dotted curve represents the best-fit  found using an exponential decay function.}
 \label{fig:1a}
\end{figure}

For both novae we reduced the UVES and X-Shooter data using a pre-compiled pipeline based on the \emph{python-cpl} libraries, which make use of the standard ESO Recipe Execution Tool (\emph{esorex}). 

\begin{table}
 \caption{Log of the observations.}
 \label{tab:1}
 \begin{tabular}{lcccc}
  \hline
  Epoch & Instrument & Exp. time & Wav. range & Resolution $R$ \\
  (Day) &  & $(s)$ & $(nm)$ & $(\lambda/\delta\lambda)$\\
  \hline
  \multicolumn{5}{c}{ASASSN-19qv}\\
  \hline
2 & Goodman & 1x500 & 400-800 & $\sim$1,850\\
16 & UVES & 3x900 \& 3x300 & 310-945 & 40,000\\
29 & X-shooter & 2x600 \& 2x300 & 300-2,500 & 6,700-8,900\\
81 & X-shooter & 2x600 \& 2x300 & 300-2,500 & 6,700-8,900\\
  \hline
  \multicolumn{5}{c}{ASASSN-20ni}\\
  \hline
4 & UVES & 1x1800 \& 3x900 & 310-945 & $\sim$1,850\\
17 & UVES & 2x600 & 380-945 & 40,000\\
20 & UVES & 1x2200 \& 2x600 & 310-945 & 40,000\\
29 & UVES & 1x2200 \& 3x600 & 310-945 & 6,700-8,900\\
40 & UVES & 1x1800 \& 2x900 & 310-945 & 6,700-8,900\\
\hline
 \end{tabular}
\end{table}

\section{Data Analysis}

\subsection{ASASSN-19qv}

The light curve of ASASSN-19qv was obtained using AAVSO \citep{Kafka2021} data and TESS public data\footnote{https://heasarc.gsfc.nasa.gov/docs/tess/}. TESS data cover the range between 600 and 1000 nm with a very high temporal sampling. At    a cadence of 0.5 hour cadence,   they reveal presence of  fluctuations with amplitude of 0.1-0.2 mag on a timescale of a few hours, see Fig. \ref{fig:1}. There is a lack of data after $\sim$ 60 days from the nova discovery. The $V$-band early evolution is well modelled as an exponential decay with constant decay $b = 0.05$ mag/days. The best-fit with such a function provides also an estimate for the $t_2$ value, which results to be $t_2 = 11.0$ days, implying that ASASSN-19qv is close  to being classified as a fast nova, according to the classification of  \citet{Payne1957}.

\begin{figure}
 \includegraphics[width=\columnwidth]{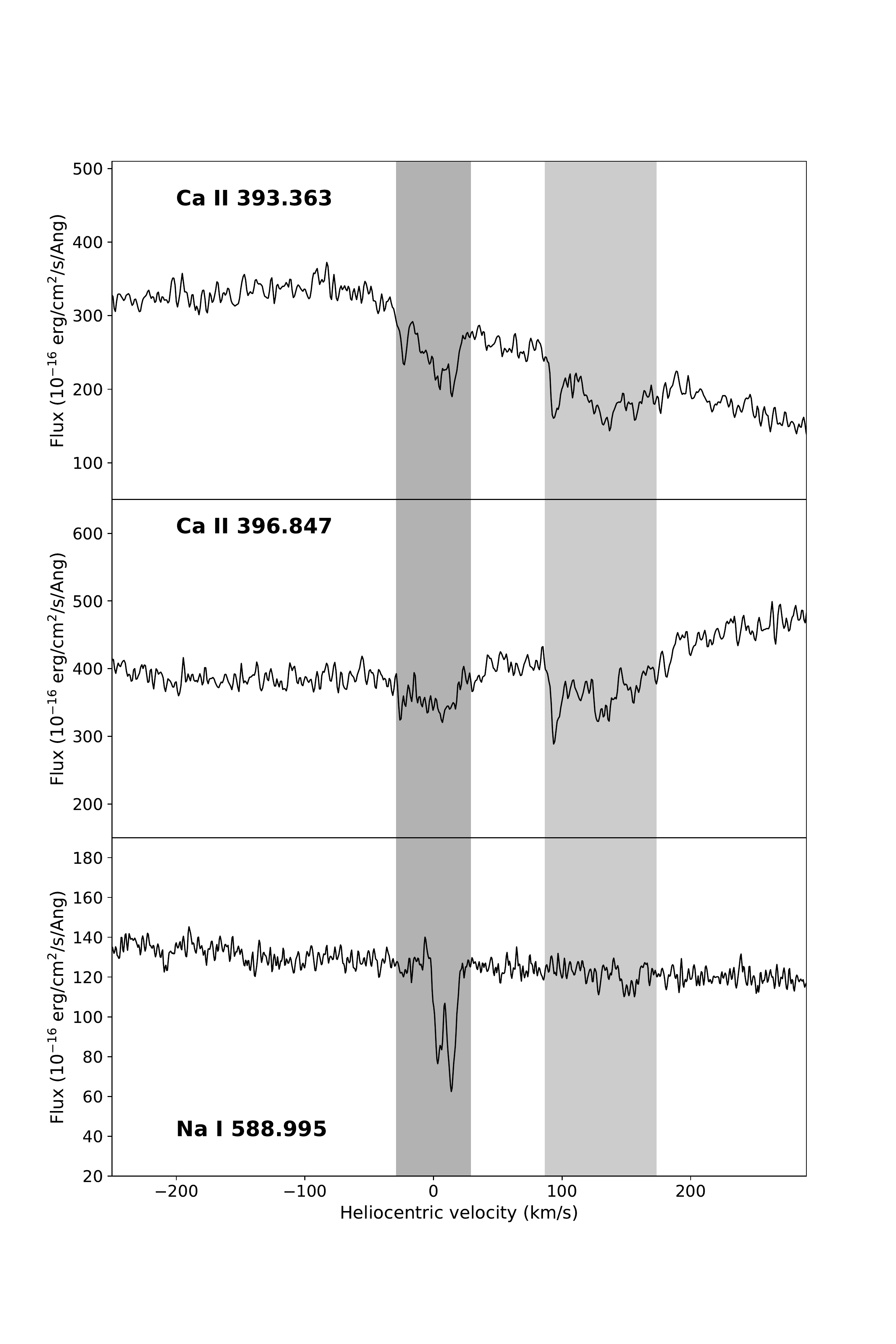}
 \caption{The Day 16 spectrum of ASASSN-19qv centered around the \ion{Ca}{ii} $\lambda$393.3 (upper panel), the \ion{Ca}{ii} $\lambda$396.8 lines (middle panel) and around the \ion{Na}{i} $\lambda$588.995 nm IS line (lower panel). Velocities are corrected for the heliocentric correction. The SMC interstellar \ion{Ca}{ii} absorptions are observed at $v_{SMC} = +130 km/s$ and are broader than the corresponding Milky Way (MW) lines. These are marked with a light gray strip, while the MW component is reported with a darker strip. }
 \label{fig:2}
\end{figure}

The high-resolution spectrum provided by UVES allows us to identify SMC interstellar lines, like  \ion{Ca}{ii} H,K lines, and then to determine the velocity offset due to the motion of the SMC that will be considered in the  analysis presented in this work. In Fig. \ref{fig:2} we show the heliocentric radial velocity of both \ion{Ca}{ii} IS lines, in addition to the \ion{Na}{i} D2 line. While \ion{Na}{i} is almost absent in the SMC environment, the \ion{Ca}{ii} lines are clearly detected and they are characterised by a main absorption centered at $v_{SMC,1} = 130$ km/s, which will be considered as the main SMC offset in the rest of the analysis for this nova. We also note a narrow component at lower velocities, $v_{SMC,2} = 90$ km/s likely due to an additional cloud of interstellar gas in the SMC along our line of sight. 

\subsubsection{The spectroscopic evolution}

\begin{figure}
 \includegraphics[width=\columnwidth]{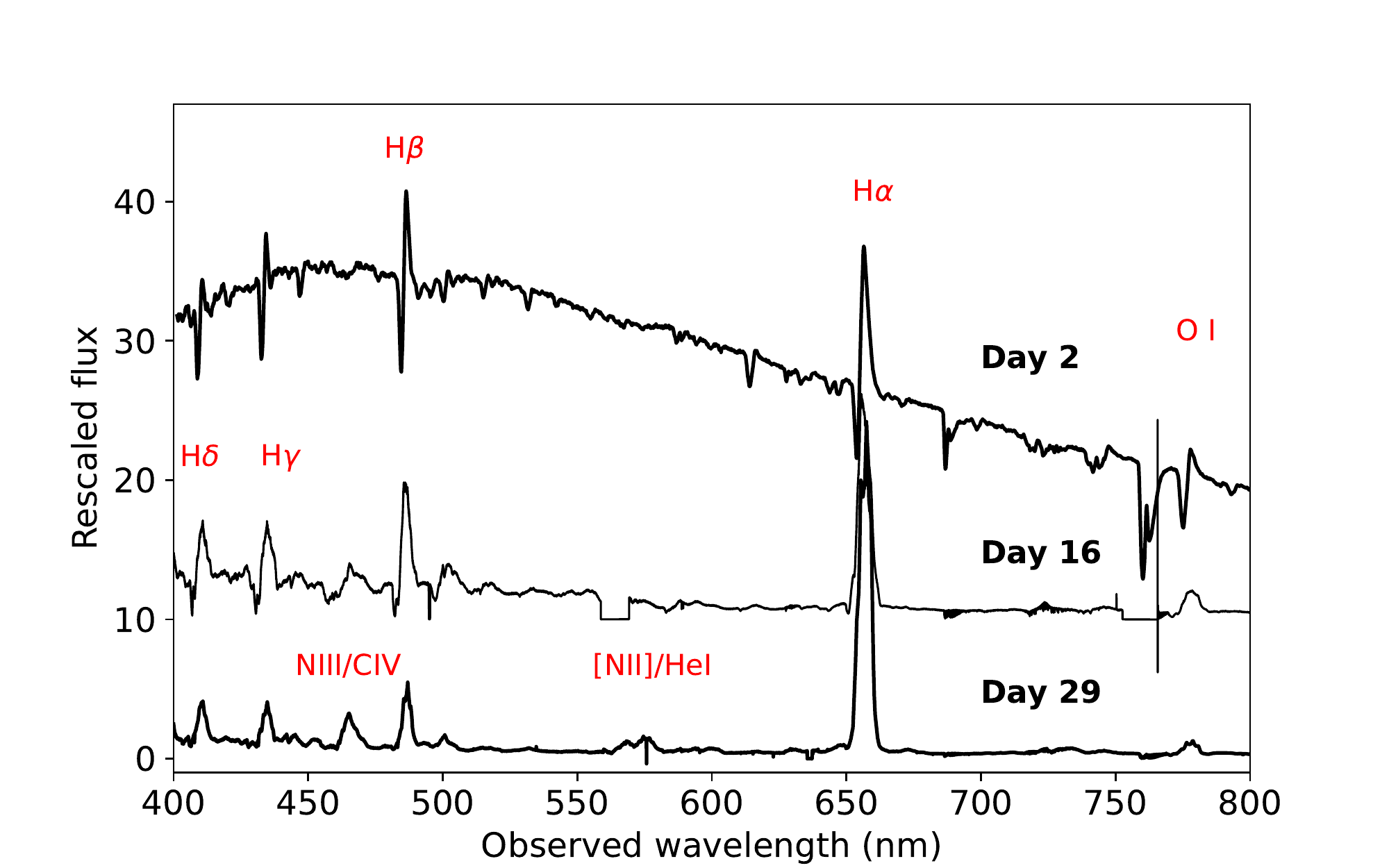}
 \caption{Spectral evolution of the nova ASASSN-19qv in the first month after the discovery. Spectra have been rescaled in flux and shown in the range (400, 800) nm. The presence of helium, blended with \ion{Fe}{ii}, nitrogen lines and of a bright Bowen blend at 464.0 nm observed in the Day 16 spectrum marks the transition from the \ion{Fe}{ii} to the He/N type, classifying this nova as a hybrid case, according to \citet{Williams1994}. In the Day 29 spectrum, the presence of a rising [\ion{N}{ii}] 575.5 nm line marks the transition to the ''auroral'' phase.}
 \label{fig:3}
\end{figure}

\begin{figure}
 \includegraphics[width=\columnwidth]{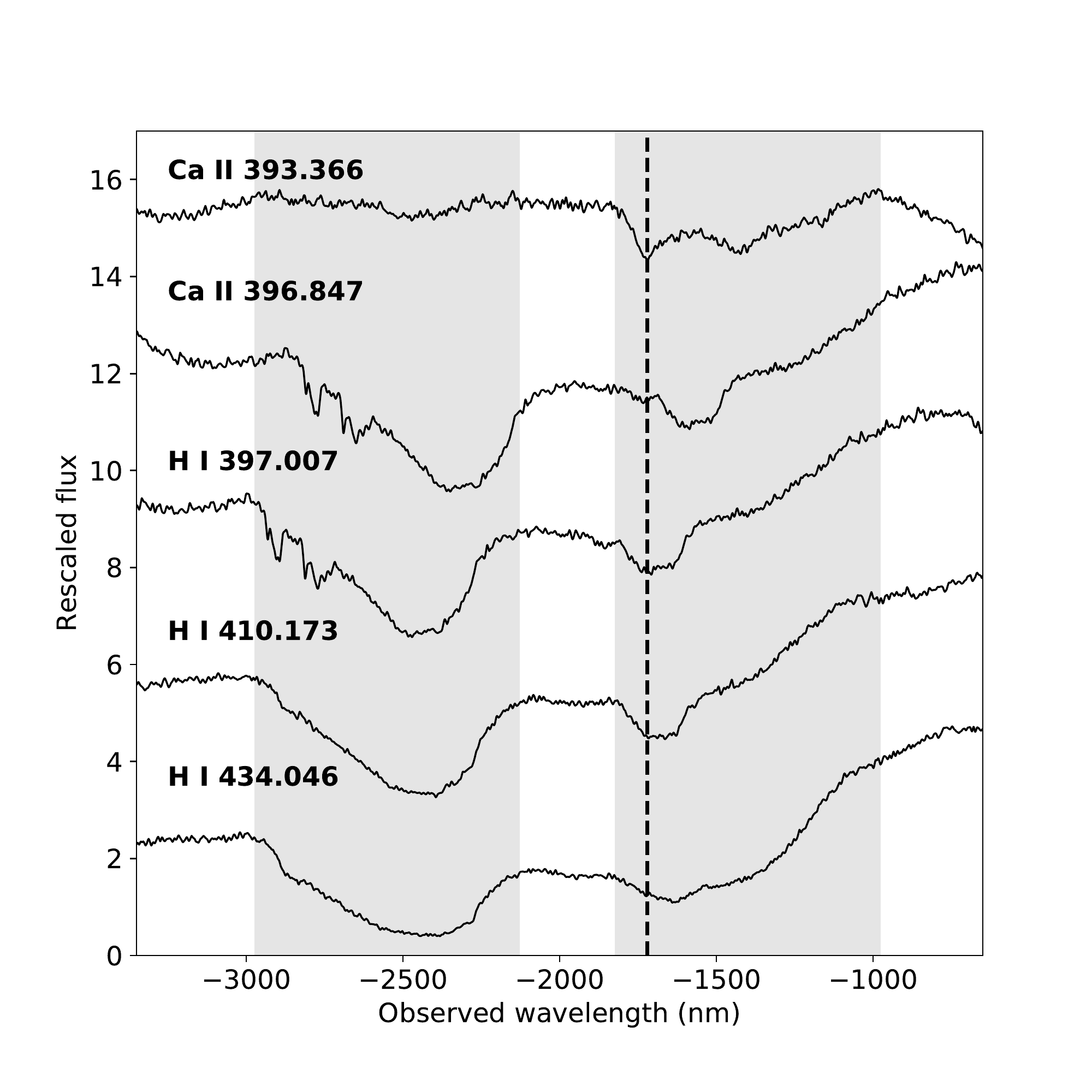}
 \caption{ Day 16 spectrum of ASASSN-19qv, corrected for the SMC motion, showing the P-Cygni absorptions of \ion{Ca}{ii} H,K lines and Balmer H$\gamma$, H$\delta$ and H$\epsilon$ lines. The absorption components are clearly visible in all Balmer lines. The \ion{Ca}{ii} 396.847 nm is blended in the broad P-Cygni of H$\epsilon$ and only the prominent feature at $v = -1720$ km/s is clearly visible (the dashed black line). The high velocity component in the \ion{Ca}{ii} 393.366 nm line is less pronounced, suggesting a very low density for the Calcium in this component.}
 \label{fig:4}
\end{figure}

Our first spectrum covers the 400\,--\,800 nm range and  was observed only two days after  discovery. It is characterised by the presence of Balmer, \ion{Na}{i} and \ion{Fe}{ii} P-Cygni absorption lines with blue-shifted absorption velocities between  $-$900 and $-$1,000 km/s. The \ion{O}{i} $\lambda$777.5 nm line is the brightest non-Balmer line, implying that this spectrum is typical of the \ion{Fe}{ii} nova spectral class, see also Fig. \ref{fig:3}.

The second spectrum obtained 16 days after the nova discovery peaks at bluer wavelengths than the first spectrum. It shows P-Cygni profiles for Balmer lines characterised by two main absorption components, at the blue-shifted velocities of $v_{1,a} = -1,710$ km/s and $v_{1,b} = -2,400$ km/s. The  brightest non-Balmer line is \ion{O}{i} $\lambda$844.6 nm. The \ion{Fe}{ii} $\lambda$516.9nm line now shows fainter P-Cygni absorptions at the same velocities of the Balmer lines, indicating that the ionization state is increasing. This evidence is also confirmed by the presence of faint, but emerging, \ion{He}{i} lines, in particular the \ion{He}{i} $\lambda$501.6 nm line that is blended with \ion{Fe}{ii} $\lambda$501.8 nm, and the presence of \ion{N}{i} and \ion{C}{ii} lines. The third spectrum (Day 29) shows higher ionization transitions like the Bowen (\ion{N}{iii}-\ion{C}{iv}) blend at 464.0 nm in addition to a rising [\ion{N}{ii}] $\lambda$575.5 nm line, which mark the beginning of the ''auroral'' phase. In the near-IR range the brightest line, with the exception for the Paschen-$\alpha$, is \ion{O}{i} $\lambda$1128.7 nm as it is expected given the high luminosity of the \ion{O}{i} $\lambda$844.6 nm line. This evidence suggests that the photo-excitation by accidental resonance \citep{KastnerBhatia1995} is still working at this epoch. The ratio $R_{\ion{O}{i}}$ between \ion{O}{i} $\lambda$844.6 nm / \ion{O}{i} $\lambda$777.5 nm has been proposed by \citet{Williams2012} as a density diagnostic for the ejecta. We measure $R_{\ion{O}{i},1} = 3.5$ in the Day 14 spectrum and $R_{\ion{O}{i},1} = 12.6$ in the Day 27 spectrum, suggesting a relatively high density of the ejecta, more similar to \ion{Fe}{ii}-type novae, but rapidly decreasing between the two epochs. 

The wavelength range of the spectra obtained with UVES and X-shooter extends to the near-UV part of the electromagnetic spectrum, permitting to analyse with a good signal-to-noise the region where the resonance transition of the $^7$\ion{Be}{ii} $\lambda$ 313.0/1 nm  doublet isotope falls. Indeed, in both early epochs (Day 14 and Day 27) we detect blue-shifted absorption components that we have identified as due to the $^7$\ion{Be}{ii} $\lambda$ 313.0/1 nm transition, see also Figs. \ref{fig:4}, \ref{fig:5}. In the Day 14 spectrum we clearly see two broad P-Cygni absorptions that share the same expanding velocities of the P-Cygni absorptions observed in Balmer lines, as well as in other transition as \ion{O}{i}. These absorption components are characterised by expanding velocities of $v_{Day16,1} = -1,630$ km/s and $v_{Day16,2} = -2,400$ km/s as observed in the Day 14 spectrum, with the slow component being less intense than the fast one. This difference is more clear in the spectrum obtained on Day 27, where the slow expanding component, with a measured velocity of $v_{Day29,1} = -1,790$ km/s, becomes more narrow while the fast component is characterised by a broad profile ($FWHM_{H\beta} \approx 900$ km/s) centered at the value of $v_{Day29,2} = -2,600$ km/s. This behavior is observed in all the main Balmer lines as well as in the $^7$\ion{Be}{ii} $\lambda$ 313.0/1 nm transition. This is the main evidence that beryllium was synthesized during the TNR preceding the outburst of ASASSN-19qv.

The Day 81 spectrum shows the absence of high-velocity blue-shifted absorption features, observed in the previous epochs, but we still detect permitted transitions like \ion{O}{i} 844.6 nm (including the 1129.1 nm in the near-IR range), and the flux ratio with the \ion{O}{i} 777.5 nm which is blended with [\ion{Ar}{iii}]  is now  $R_{\ion{O}{i}} = 8.1$, suggesting an environment still relatively dense. There are also high-ionization lines such as \ion{He}{i} ground state transitions 1,008.3/2,005.8 nm as well as excited transitions (587.6/706.5 nm lines among the many others barely discernible in the spectrum). We also detect the presence of typical forbidden transitions observed in the nebular phase of classical novae like the [\ion{O}{iii}] 436.3/495.9/500.7 nm, the [\ion{N}{ii}] 575.5 nm, suggesting that the physical conditions of the ejecta at this stage are very heterogeneous (see Fig. \ref{fig:7}). Then we can not use this spectrum to infer physical properties like the density, temperature and the hydrogen mass of the ejecta. We do not detect forbidden neon lines at this epoch, suggesting a CO-type WD for the nova progenitor.

\subsection{ASASSN-20ni}

The light curve of ASASSN-20ni was built using AAVSO data \citep{Kafka2021}. We have also used ASAS-SN $g$-band data that covers the rising phase of the nova and its decay up to $\sim$ 60 days after the peak brightness, see Fig. \ref{fig:1a}. The last non-detection from the ASAS-SN survey was dated October  25, 2020, e.g. one day before the first nova detection by the same project \citep{Way2020}. The light curve evolution in the first 60 days of the nova emission can be modelled using an exponential function, despite the light curve showing a short rebrightening at $\sim$ 20 days from the peak brightness. Using the AAVSO $V$-band data, we measure a decay rate of $b = 0.02$ mag/days, and a $t_2$ parameter of $t_2 = 17.4$ days, which classify ASASSN-20ni as a fast nova.  

\begin{figure}
 \includegraphics[width=\columnwidth]{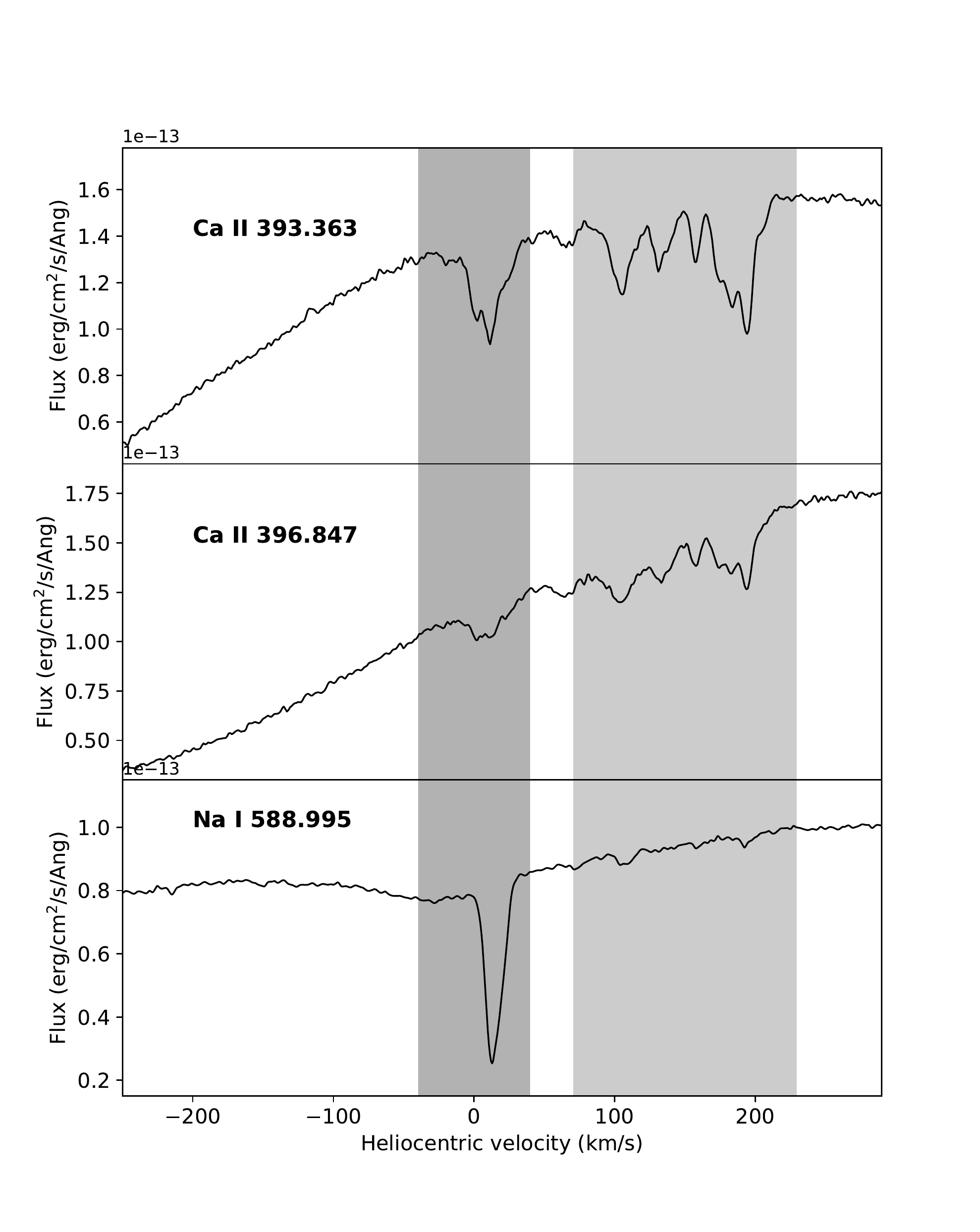}
 \caption{The Day 4 spectrum of ASASSN-20ni centered around the \ion{Ca}{ii} $\lambda$393.3 (upper panel), the \ion{Ca}{ii} $\lambda$396.8 lines (middle panel) and around the \ion{Na}{i} $\lambda$588.995 nm IS line (lower panel). Velocities are corrected for the heliocentric correction. The SMC interstellar \ion{Ca}{ii} absorptions are observed at multiple velocity (see text).}
 \label{fig:2b}
\end{figure}

ASASSN-20ni was observed in the central region of the SMC, whereas ASASSN-19qv was located at a more peripheral region, see Fig. \ref{fig:8}. This is reflected on the amount of  gas surrounding the nova location as inferred from the analysis of interstellar lines. As observed in the \ion{Ca}{II} H,K lines, we note multiple components at $v_{SMC}$ = 105, 131, 156 and 194 km/s, with the latter being the most intense component, see also Fig. \ref{fig:2b}. Similar to the spectrum of ASASSN-19qv, the \ion{Na}{I} inter-stellar component is less pronounced, showing only the components at $v_{SMC}$ = 105 and 194 km/s. In the following, we will consider the highest velocity component at $v_{SMC}$ = 194 km/s as our reference to correct all our spectral series to the SMC velocity.

\subsubsection{The spectroscopic evolution}

\begin{figure}
 \includegraphics[width=\columnwidth]{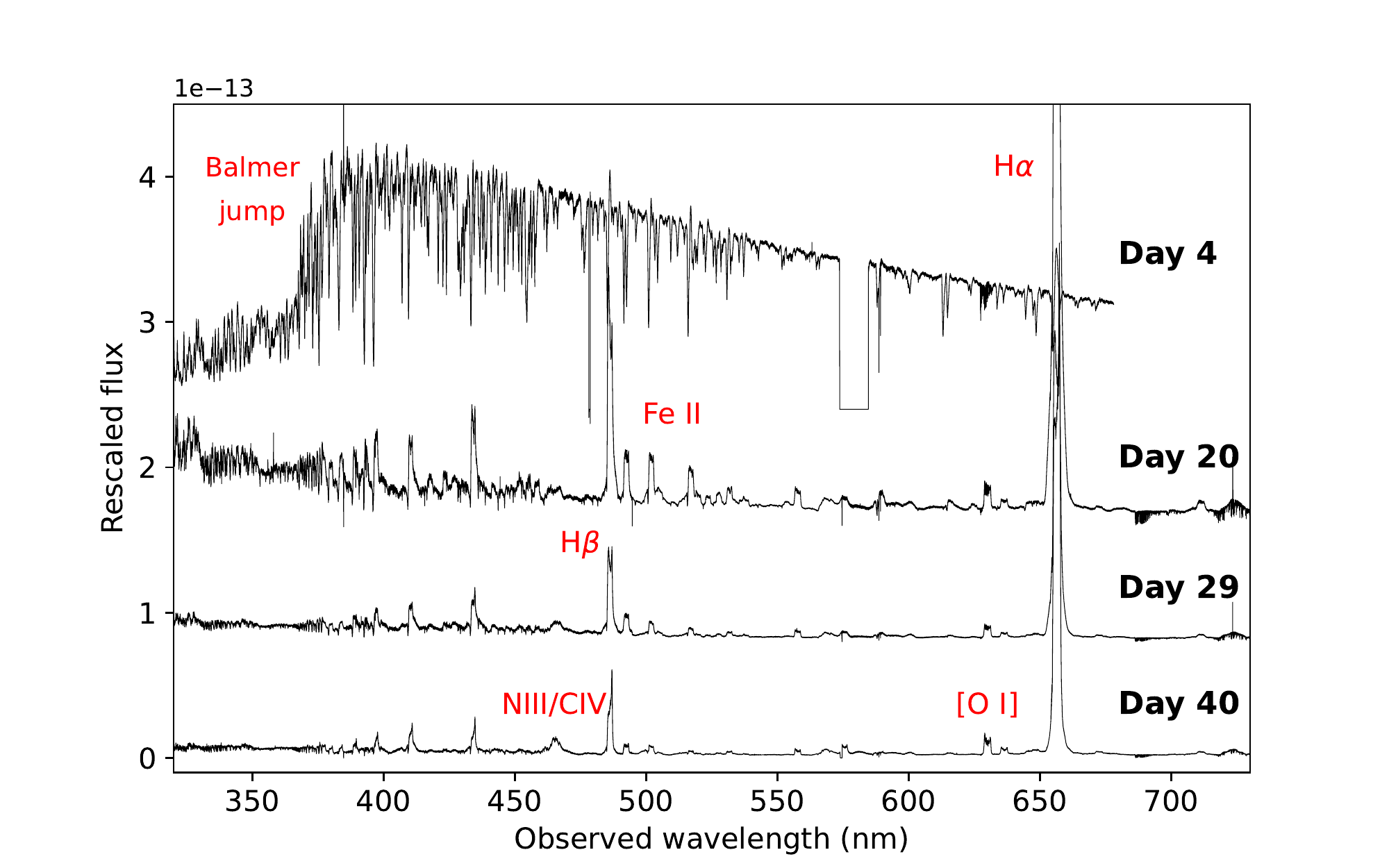}
 \caption{Spectral evolution of the nova ASASSN-19qv in the first 40 days after the discovery. Spectra have been rescaled in flux and shown in the range (330, 740) nm.}
 \label{fig:3b}
\end{figure}

The first spectrum of ASASSN-20ni was obtained just four days after its discovery. It shows a very optically thick continuum, typical of the \ion{Fe}{II} spectral class \citep{Williams1991a}, and is  characterised by the presence of a ''forest'' of absorption lines in the range between 300 and 550 nm, see also Fig. \ref{fig:3b}. After re-scaling the spectrum at the SMC velocity assumed for this nova, we measure blue-shifted absorption velocities of $v_{exp}$ $\sim$ -520 km/s for  \ion{Na}{I}, \ion{Ca}{II}, and \ion{Fe}{II} lines, while Balmer lines show a broader absorption trough extending to higher velocities. We also detect at the same expanding velocity the majority of THEA lines listed in Table 2 of \citet{Williams2008}, with the exception of \ion{V}{II} lines. We cannot clearly confirm the presence of \ion{Li}{I} 670.7nm, despite  broad absorption being observed at the corresponding blue-shifted wavelength.

\begin{figure}
 \includegraphics[width=\columnwidth]{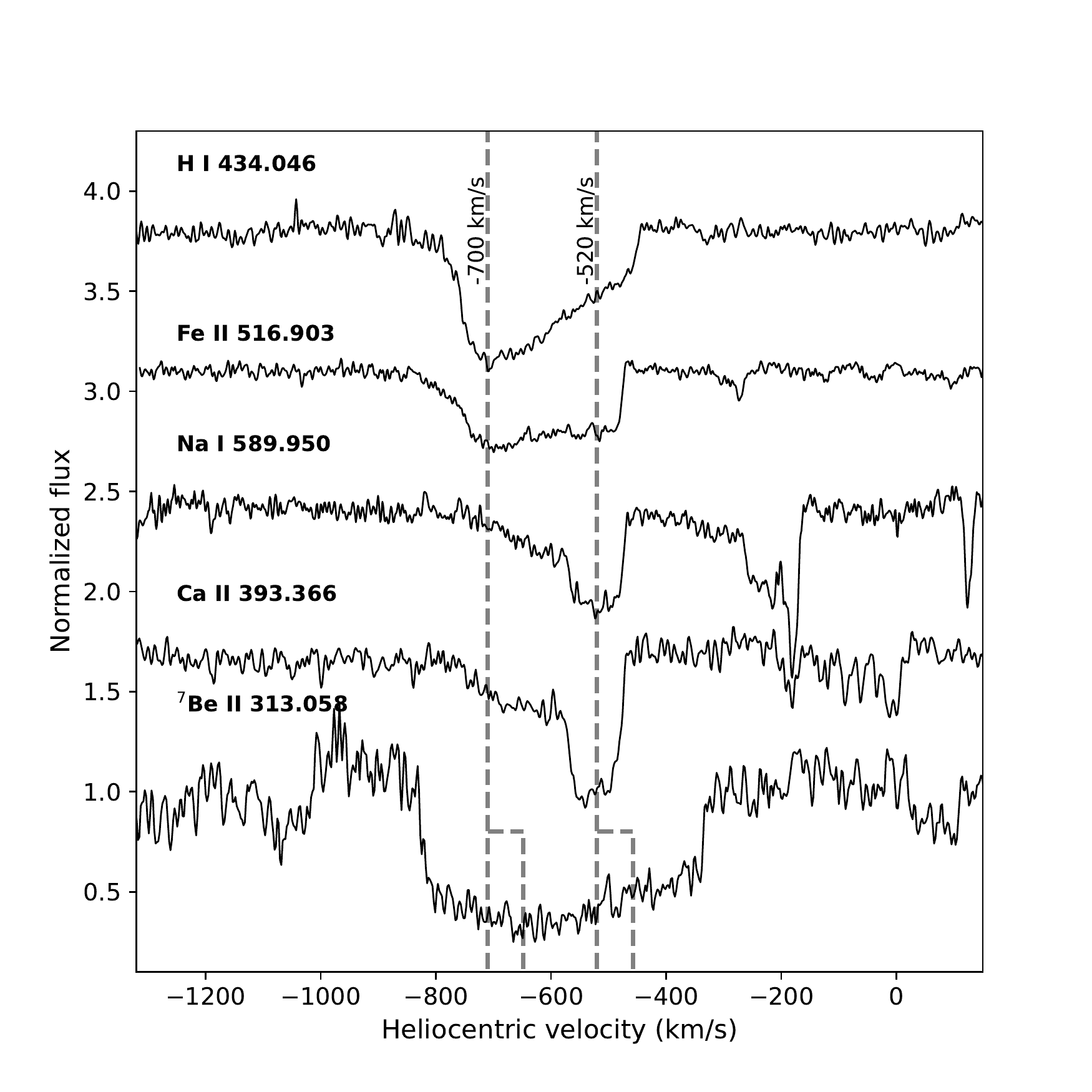}
 \caption{The Day 40 spectrum of ASASSN-20ni, corrected for the SMC motion, showing the P-Cygni absorptions of \ion{Ca}{ii} K, H$\gamma$, \ion{Fe}{II} 516.9 nm, \ion{Na}{I} 589.9 nm and of $^7$\ion{Be}{II} 313.0 nm lines. The plot shows the different profile of \ion{Ca}{II} and \ion{Na}{I}, which show a prominent absorption at -530 km/s and a more faint high-velocity tail, with respect to \ion{H}{I}, \ion{Fe}{II} and $^7$\ion{Be}{II} lines where the higher velocity component centered at $\sim$ -700 km/s and extending to -800 km/s is more pronounced. This evidence suggests a different abundance composition for the two components and supports a two-ejecta component scenario \citep{Mukai2019}. for the $^7$\ion{Be}{II} line are reported the expected positions of the two doublet components.}
 \label{fig:3c}
\end{figure}

The second spectrum, obtained two weeks later (Day 17), and covering the range from 375 to 500 nm and from 580 to 946 nm, shows an evolved spectrum where almost all THEA lines have disappeared.  The brightest non-Balmer line is \ion{O}{I} 844.6 nm, showing a main P-Cygni absorption system at higher velocities $v_{exp}$ = -700 km/s, similar to what is measured for Balmer and \ion{Fe}{II} lines. Interestingly, the \ion{Na}{I} doublet shows a more structured profile, with signatures of two components at velocities of $v_1 = -650$ km/s and $v_2 = -520$ km/s. This velocity configuration for the ejecta is also observed in the following spectra (Day 20 and Day 29). At these epochs, bright emission lines display a saddle-shaped profile with the two bright peaks at $\pm \sim$ 500 km/s, which suggests an asymmetric geometry for the nova ejecta \citep{Mukai2019}. The last spectrum obtained on Day 40 still presents low-ionization transitions such as the \ion{Fe}{II} multiplet 42, but we also see enhanced higher-ionization transitions such as  the Bowen blend in emission and the \ion{He}{I} 587.5 nm, which was identified by the presence of a P-Cygni absorption with blue-shifted velocity of $\sim$ -650 km/s. Interestingly, lower ionization transitions such as \ion{Ca}{II}, \ion{Fe}{II} and \ion{Na}{I} show a more intense lower velocity ($\sim$ -520 km/s) component over-imposed to the higher velocity component with  intensity fading with the increasing velocity to -700 km/s, see Fig. \ref{fig:3c}.

\section{Abundance of $^7$Be}\label{sec:212}

\subsection{ASASSN-19qv}
\begin{figure}
 \includegraphics[width=\columnwidth]{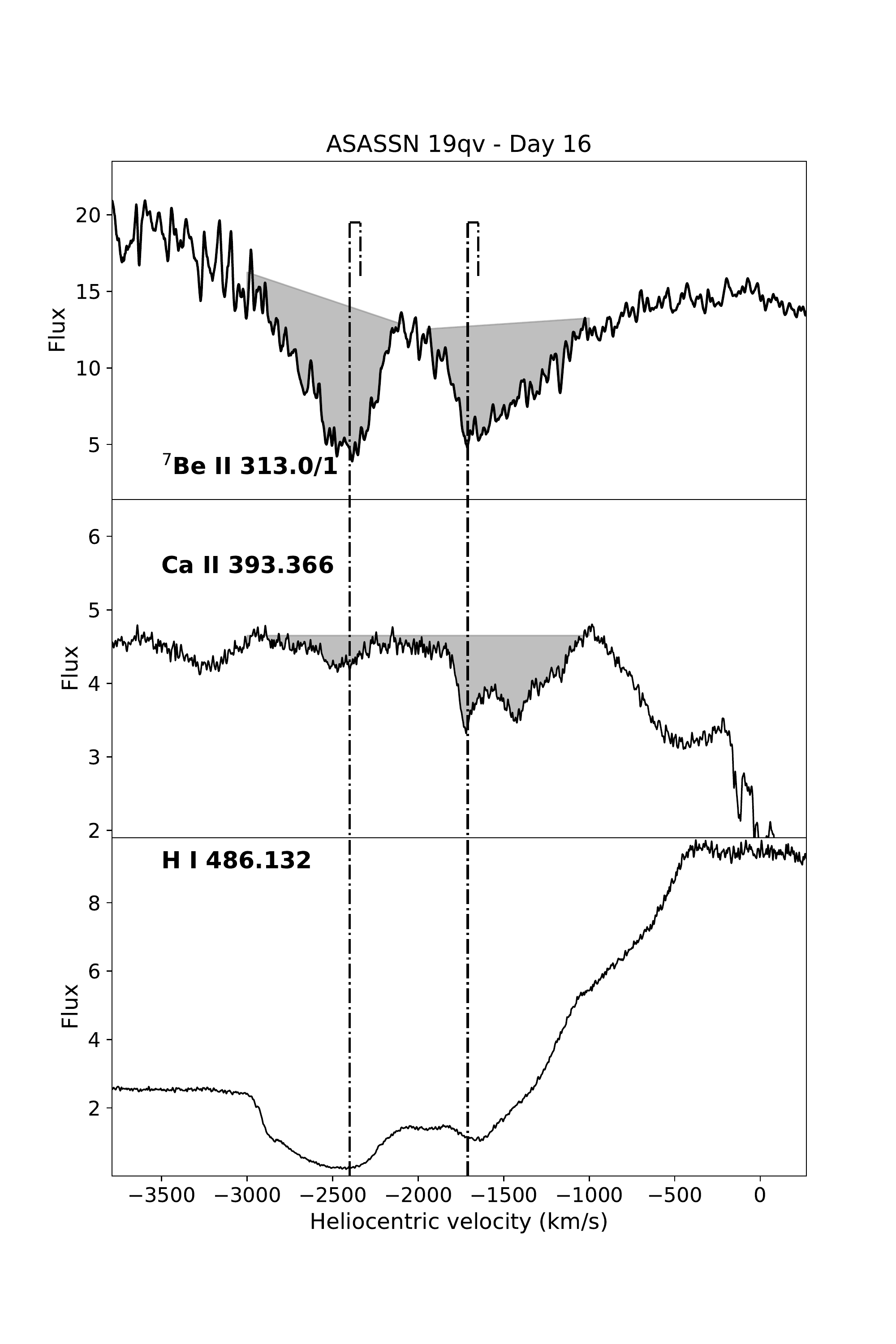}
 \caption{The Day 16 spectrum of ASASSN-19qv, corrected for the SMC motion, showing the P-Cygni absorptions of the $^7$\ion{Be}{ii} $\lambda$313.0 at blue-shifted velocities of $v_{1a}$ = -1710 km/s and $v_{1b}$ = -2400 km/s and compared with the \ion{Ca}{ii} $\lambda$396.8 and with H$\beta$ . The dashed regions mark the area of the absorptions considered for the EW estimate.}
 \label{fig:5}
\end{figure}

\begin{figure}
 \includegraphics[width=\columnwidth]{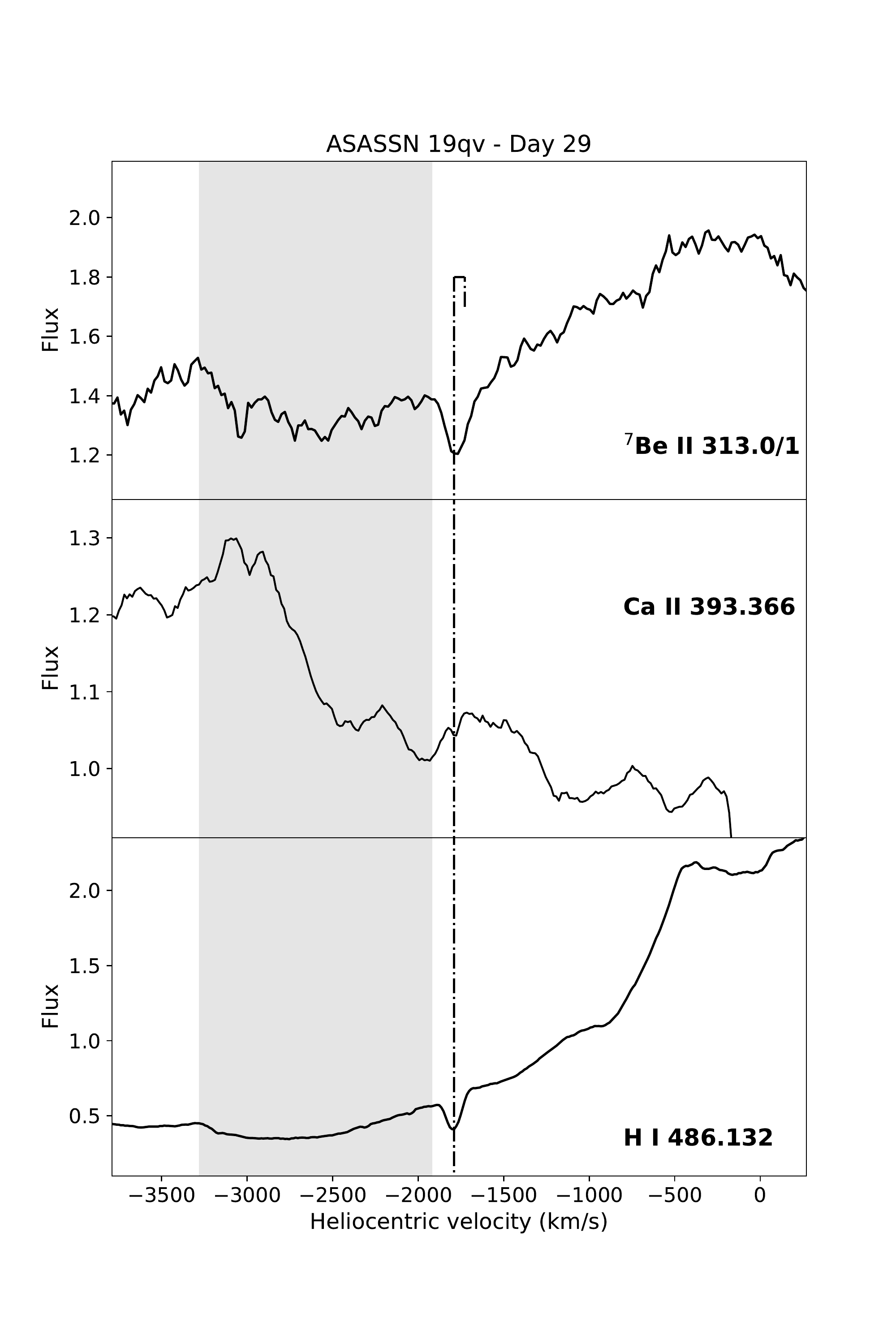}
 \caption{The Day 29 spectrum of ASASSN-19qv, corrected for the SMC motion, showing the P-Cygni absorptions of the $^7$\ion{Be}{ii} $\lambda$313.0, the \ion{Ca}{ii} $\lambda$396.8 and with H$\beta$ at blue-shifted velocities of $v_{1a}$ = -1710 km/s, while the gray dashed region indicates the broad component centered at $v_{1b}$ = -2600 km/s.}
 \label{fig:6}
\end{figure}

\begin{figure*}
 \includegraphics[width=2.0\columnwidth]{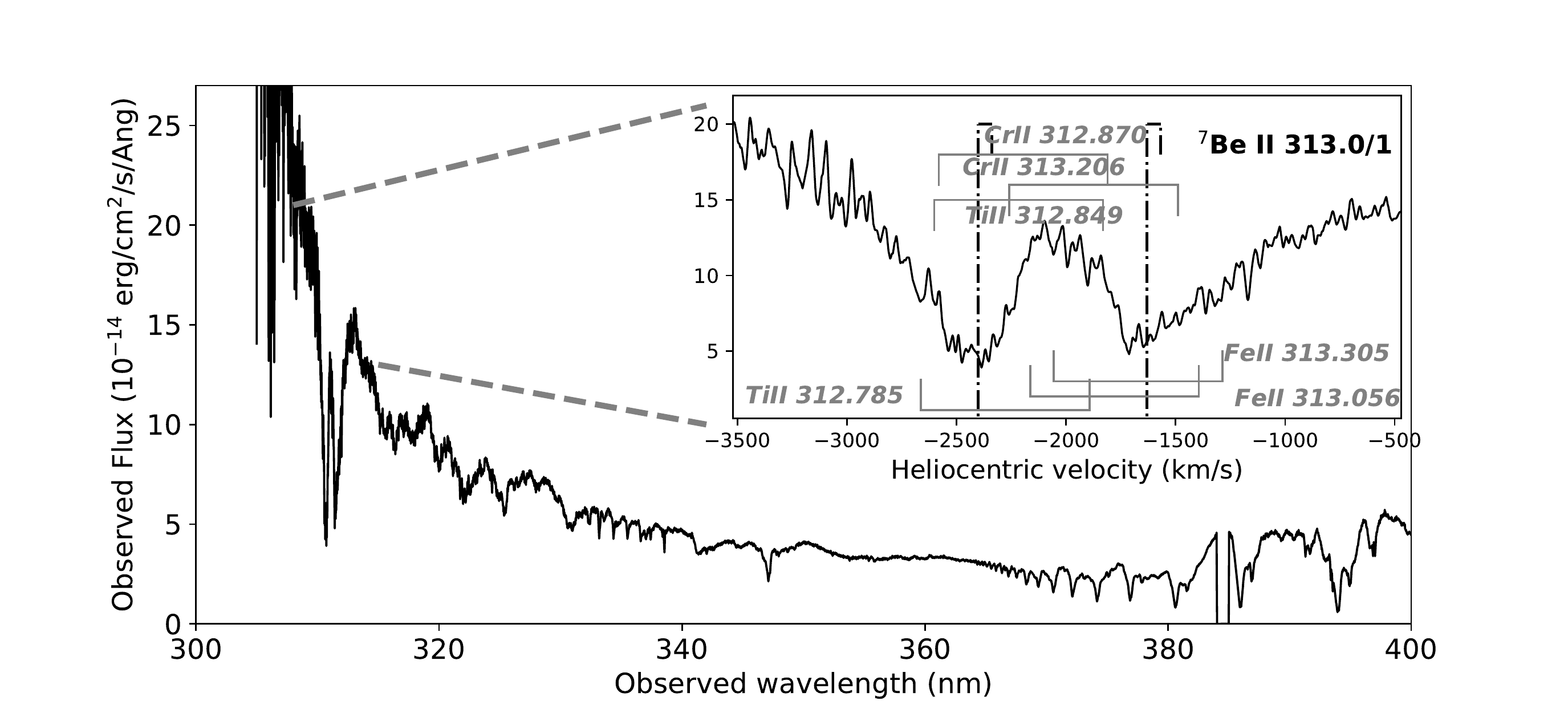}
 \caption{The Day 16 spectrum of ASASSN-19qv in the region 300-400 nm. The inset plot shows the region centered around the \ion{Be}{ii} $\lambda$313.058, which shows the expected positions of low ionisation transition absorptions of \ion{Cr}{ii}, \ion{Ti}{ii} and \ion{Fe}{ii}. Velocities are corrected for the heliocentric correction. }
 \label{fig:7}
\end{figure*}

Following \citet{Tajitsu2015}  we  quantified the amount of beryllium ejected in the ASASSN-19qv outburst by comparing the observed equivalent widths (EW) of the blue-shifted absorption components with the EW of a reference element.  The resonance lines of ionised calcium  \ion{Ca}{ii} $\lambda\lambda$393.4/396.8 nm H,K  are the more suitable, given that calcium shares the same configuration of their outermost electron shells. 
 A detailed analysis of  Day 16 spectrum shows the presence of faint blue-shifted components for the \ion{Ca}{ii} 393.4 nm line  centered at the same velocities observed for the Balmer lines, see Fig. \ref{fig:4}. 

The other doublet component, \ion{Ca}{ii} 396.8 nm, is blended within the P-Cygni of the \ion{H}{i} 397.0 nm line, but we confirm its presence through the identification of a faint feature corresponding to the main absorption observed for all transitions at $v = -1,720$ km/s. Moreover, we observe that the high-velocity absorption components of Calcium lines are much fainter when compared with the slower components, while for $^7$\ion{Be}{ii} and Balmer lines this is not observed. This evidence suggests that the high-velocity component ejecta has a distinct element abundance. This difference can be explained in the case that high-velocity components arise from a distinct ejecta event, like the one proposed in \citet{Mukai2019}, where a later fast wind-like, and likely bi-polar ejection is observed after the first phenomenon, which is slower and characterized by an oblated geometrical distribution. We will come back to this point later. In the spectrum obtained on Day 29, \ion{Ca}{ii} lines almost disappeared.  In the Day 29 spectrum we observed a very faint absorption for \ion{Ca}{ii} 393.3 nm line, see Fig. \ref{fig:5}, and in order to estimate the beryllium abundance we consider  the analysis of the UVES Day 16 spectrum. 

\begin{table}
\caption{A list of single-ionized ions which are the main contributors  to
the absorption in the region around  $\lambda$313.058 and that are shown in Fig. \ref{fig:7}.  Atomic data  taken from the  NIST lines data base. Those for \ion{Cr}{II} are from \citet{Lawler2017}. }
\label{tab:3}
\begin{center}
\begin{tabular}{crccc}
\hline
\hline
Wavelength (Air) & ion   & Log(gf) & Low en. & Upper en.  \\
nm &  &  &   (eV) & (eV) \\
\hline
\hline
   312.7850 & Ti {\sc ii} &        0.15  & 3.87 & 7.83 \\
   312.8483 & Ti {\sc ii} &        0.11  & 3.90 & 7.87 \\
   312.8700 & Cr {\sc ii} &        -0.53  & 2.43 & 6.40 \\
   313.0565 & Fe {\sc ii} &          --     & 3.77 & 7.73   \\
   313.2057 & Cr {\sc ii} &      0.43   & 2.48 & 6.44 \\
   313.3048 & Fe  {\sc ii}&        -1.9   & 3.89 & 7.84 \\
\hline
\end{tabular}
\end{center}
\end{table}

We have then measured the EWs for the $^7$\ion{Be}{ii} and the \ion{Ca}{ii} 393.4 nm for the blue-shifted absorption area as shown in Fig. \ref{fig:5}. However, as also reported in our previous analysis \citep{Molaro2020}, the absorption of $^7$\ion{Be}{ii} is generally contaminated by the presence of low ionisation transitions of \ion{Cr}{ii}, \ion{Ti}{ii} and \ion{Fe}{ii}. We have then analysed the presence of the most intense transitions (see Table \ref{tab:3}) that fall close to the wavelength range of $^7$\ion{Be}{ii} and in Fig. \ref{fig:7} we report  their expected position in the spectral region surrounding the blue-shifted beryllium absorptions. We cannot clearly confirm the presence of \ion{Cr}{ii} and \ion{Ti}{ii}, given the relatively low signal-to-noise ($\sim$10) of the UVES spectrum around the $^7$\ion{Be}{ii} transitions. We have also checked for the presence of absorption lines from the other transitions of these same elements originating from their parent multiplets, but it is hard to firmly establish their presence. We find possible evidence of  the presence of \ion{Fe}{ii} lines $\lambda\lambda$313.305,313.056, and other transitions arising from the same initial level like \ion{Fe}{ii} $\lambda$316.794,314.472, whose detection, in addition to the already confirmed presence of \ion{Fe}{ii} $\lambda$516.903, confirms the presence of iron in the ejecta of ASASSN-19qv. Consequently, in our final estimates of the $^7$\ion{Be}{ii} EWs,  we are taking into account the presence of these line blends.

Our final results are shown in Table \ref{tab:2}, where we report the measured EWs for the low and high velocity components, as well as the total value. The EW ratio between $^7$Be and Ca for low velocity component, EW($^7$\ion{Be}{ii}$_{low}$)/EW(\ion{Ca}{ii}$_{low}$) = 2.26 $\pm$ 0.07,  while for the high velocity component we measure a much higher value, namely EW($^7$\ion{Be}{ii}$_{high}$)/EW(\ion{Ca}{ii}$_{high}$) = 8.58$\pm$0.07. In the following analysis we will not distinguish between these two emission components, given that both systems are escaping the binary and will then enrich the ISM of the SMC.  But, a detailed analysis included in a wider context is needed and will be presented elsewhere. For the entire absorption systems we consider the average ratio between the two components, obtaining EW($^7$\ion{Be}{ii})/EW(\ion{Ca}{ii}) = 3.50$\pm$0.09. This value slightly exceeds the mean value of $\sim 1.5 \pm 0.2$ found in Galactic  novae \citep{Molaro2020}. Following \citet{SpitzerBook}, the relative number abundance can be inferred from:
\begin{equation}
    \frac{N(^7\ion{Be}{ii})}{N(\ion{Ca}{ii})} = 2.164 \times \frac{EW(^7\ion{Be}{ii})}{EW(\ion{Ca}{ii})}
\end{equation}
$^7$\ion{Be}{ii} is unstable with an  half-life time decay of $t_{1/2} = 53.3$ days. Thus, considering the correction factor of $K = 1.23$  we obtain N($^7$\ion{Be}{ii}) = (9.32$\pm$0.30) $\times$ N(\ion{Ca}{ii}). 

\begin{table*}
\centering
\caption{Equivalent width measurements for the $^7$\ion{Be}{ii} and \ion{Ca}{ii} 393.4 nm blue-shifted absorptions in the Day 16 spectrum of ASASSN-19qv.}
\label{tab:2}
\begin{tabular}{lccccccc}
\hline \hline
   Day    & EW($^7$\ion{Be}{ii}$_{low}$) &  EW($^7$\ion{Be}{ii}$_{high}$) & EW($^7$\ion{Be}{ii}$_{tot}$)  & EW(\ion{Ca}{ii}$_{low}$) & EW(\ion{Ca}{ii}$_{high}$) & EW(\ion{Ca}{ii}$_{tot}$)\\
       & (\AA) & (\AA) & (\AA) & (\AA) & (\AA) & (\AA)     \\
    \hline
16 & 3.32$\pm$0.05 & 3.09$\pm$0.07 & 6.41$\pm$0.09 & 1.47$\pm$0.01 & 0.36$\pm$0.02 & 1.83$\pm$0.02 \\
\hline
        \end{tabular}
\end{table*}

\subsection{ASASSN-20ni}

\begin{figure}
 \includegraphics[width=\columnwidth]{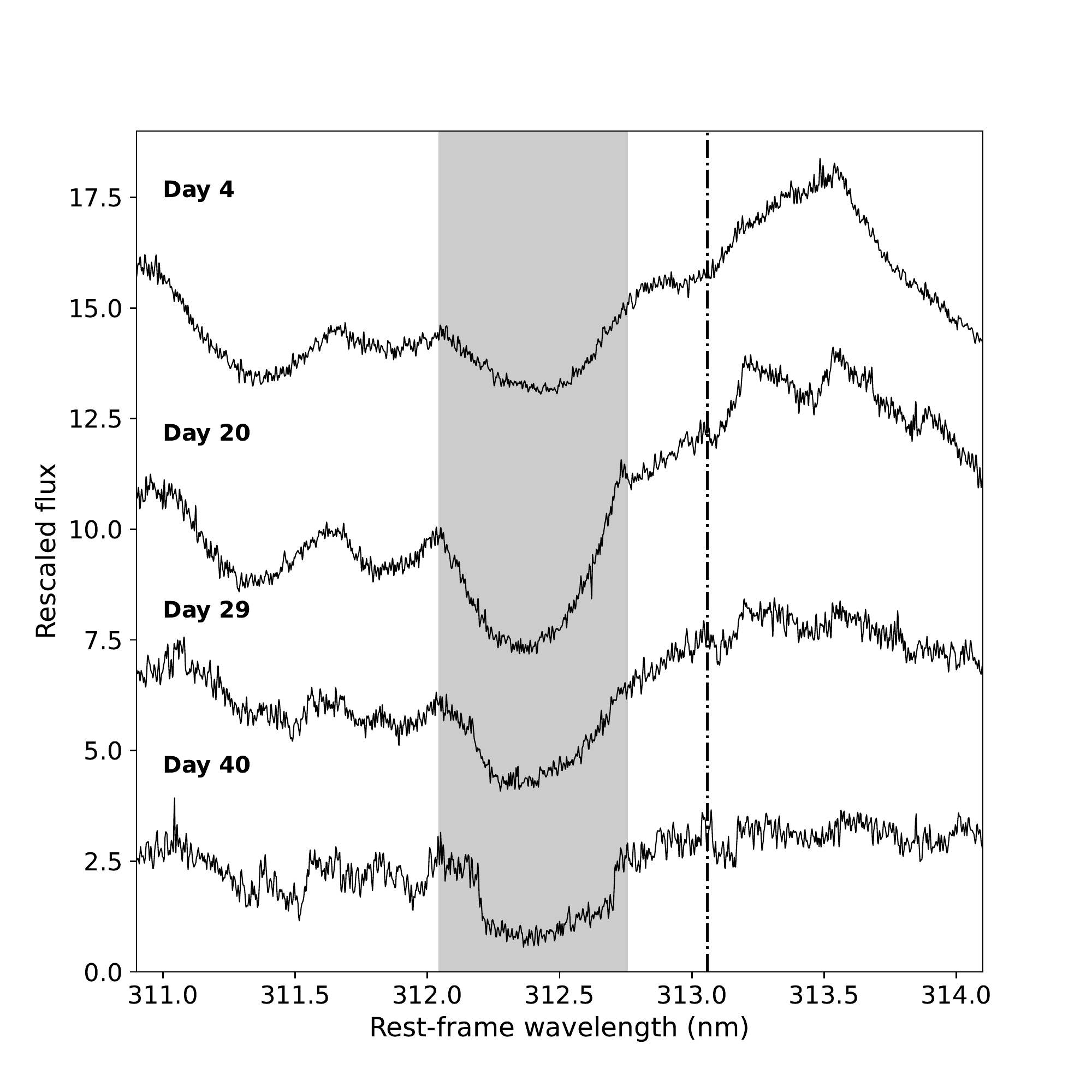}
 \caption{The evolution of the broad absorption feature centered at $\lambda$ = 312.5 nm, marked with a gray shaded region, and attributed to $^7$\ion{Be}{II} 313.0/1 nm doublet. The dot-dashed black line marks the rest-frame position of the $^7$\ion{Be}{II} 313.0 nm line.}
 \label{fig:9}
\end{figure}

The velocity distribution of the ejecta along the evolution of the outburst is crucial to identify the presence of features due to the doublet resonance lines of $^7$\ion{Be}{II}. In Fig. \ref{fig:9} the evolution of the region between 310 and 314 nm  is shown  to highlight the presence of a broad absorption around $\sim$ 312.5 nm. This absorption   corresponds to $^7$\ion{Be}{II} 313.0/1 nm and  matches the blue-shifted velocities reported for  other ions observed in the spectrum at the same epoch. 

Unfortunately, our spectral observations cover the first 40 days only, such that we could not follow-up the evolution of the $^7$Be P-Cygni profile until narrow features would appear, as due to the expanding ejecta and its consequent density decay, similar to the case of ASASSN-19qv  previously discussed. Moreover, in ASASSN-20ni we could not disentangle the low and high velocity components, being apparently embedded in the observed absorption profile, see Fig. \ref{fig:3c}. Consequently, we have done a careful measurement of the EW of the total absorption profile, paying attention to the contribution of other lines, such as \ion{Fe}{II}, \ion{Ti}{II} and \ion{Cr}{II} (see the list on Table \ref{tab:3}), similarly to what was done for ASASSN-19qv. The presence of THEA lines in the early spectra of ASASSN-20ni represents important information for our analysis: we already know their expanding velocities  which corresponds to the lower velocity component, and then their position in the spectrum. 

In Table \ref{tab:4} we report on our final measurement of the EW of the $^7$\ion{Be}{II} 313.0/1 nm blue-shifted absorption, as well as of \ion{Ca}{II} 393.3 nm, which we will use as our reference for the estimate of the Be abundance, as already discussed in Section \ref{sec:212}. The EW ratio between $^7$Be and Ca is EW($^7$\ion{Be}{ii})/EW(\ion{Ca}{ii}) = 2.14$\pm$0.24, as estimated from the Day 40 spectrum. Assuming a correction factor of $K = 2.11$ for the $^7$\ion{Be}{II} decay, finally we obtain 
N($^7$\ion{Be}{ii}) = (9.81$\pm$0.52) $\times$ N(\ion{Ca}{ii}), which is in good agreement with the Be abundance value estimated for ASASSN-19qv. 

\begin{table}
\centering
\caption{Equivalent width measurements for the $^7$\ion{Be}{ii} and \ion{Ca}{ii} 393.4 nm blue-shifted absorptions in the Day 20, 29 and 40 of ASASSN-20ni.}
\label{tab:4}
\begin{tabular}{lcc}
\hline \hline
   Day    & EW($^7$\ion{Be}{ii}$_{tot}$)  & EW(\ion{Ca}{ii}$_{tot}$)\\
       & (\AA) & (\AA)     \\
    \hline
20 & 2.45 $\pm$ 0.11 & 2.58 $\pm$ 0.11 \\
29 & 2.19 $\pm$ 0.12 & 2.09 $\pm$ 0.13 \\
40 & 2.77 $\pm$ 0.18 & 1.29 $\pm$ 0.12 \\
\hline
        \end{tabular}
\end{table}

\section{Discussion}

\subsection{ $^7$Be (=$^7$Li) yields}

From our previous analysis we  infer an average beryllium abundance of N($^7$\ion{Be}{ii}) = (9.63$\pm$0.50) $\times$ N(\ion{Ca}{ii}). Following similar studies \citep{Molaro2022MNRAS.509.3258M}, we  assume that singly ionized ions of \ion{Be}{II} and \ion{Ca}{II} represent the main ionization stage for the ejecta and calcium is not produced in the nova explosion.
Calcium  can be also synthesised in massive oxygen-neon WDs where the peak temperatures can reach very high values ($T \sim 5 \times 10^8$ K and more). Some observations have indeed reported Ca abundance values in nova ejecta up to one order of magnitude larger than the Solar value \citep{Andrea1994} suggesting a possible production channel for heavy ($A \sim 40$) elements in very hot WDs \citep{Christian2018,Setoodehnia2018}.  
The lack of neon in ASASSN-19qv suggests that the progenitor WD of ASASSN-19qv is not an extremely, hot massive WD. For  ASASSN-20ni, unfortunately, due to the lack of late-time spectra, we could not verify the presence of bright forbidden neon lines in the near-UV, which would imply a ONe underlying WD. However, the larger $t_2$ value suggests a similar, if not lower, massive WD progenitor. On the other hand, \citet{Starrfield2020} found that a large over-production of $^{40}$Ca (up to ten times the Solar value) is obtained from their 1D hydrodynamic simulations of the TNR in CO novae. An increase in $^{40}$Ca abundance would lead to a corresponding increase on the $^7$Be yield, with important consequences for the lithium enrichment of the SMC. However, measuring $^{40}Ca$ abundance in nova ejecta requires a high-cadence spectral coverage of the optically thick phases, when the ejecta is relatively cold, given the strong sensitivity of calcium ionization to the ambient temperature \citep{Chugai2020,Molaro2022MNRAS.509.3258M}. We therefore assume here that  Ca  in the  ejecta  shares  the average value of the  SMC stellar populations. 
 
The mean stellar metallicity of the SMC is  $[Fe/H] = -0.59 \pm 0.06$,  obtained from the analysis of massive and young  OB stars \citep{Trundle2007,Bouret2003,Korn2000}. Recent measures from a large-scale photometric analysis of the SMC have reported an average value of $[Fe/H] = -0.95 \pm 0.08$, using the slope of the Red Giant branch as an indicator of the metallicity \citep{Choudhury2018}. The location of ASASSN-19qv is, however, quite peripheral, while ASASSN-20ni is in the  inner region of the SMC. \citet{Carrera2008} report a significant  gradient with the metallicity decreasing toward the  external regions. The inner part of the SMC is characterised by higher metallicity values ($[Fe/H]\sim -0.6$). The existence of two peaks at $[Fe/H] = -0.9/-1.0$ and at $[Fe/H] = -0.6$ in the metallicity distribution of the SMC was also reported in \citet{Mucciarelli2014}. In the following, we will use the value of $[Fe/H] = -0.6$ for both novae.
Given that the solar abundance of calcium is N(Ca)/N(H)$_{\odot}$ = 2.2 $\times$ 10$^{-6}$ \citep{Lodders2009}, we obtain that the Ca abundance value in the SMC is N(Ca)/N(H)$_{SMC}$ = (5.5 $\pm$ 0.4) $\times$ 10$^{-7}$. With this value we obtain a $^7$Be, or  $^7$Li yield of N($^7$Li)/N(H) = (5.3 $\pm$ 0.2) $\times$ 10$^{-6}$. 
 
 We are not able to estimate  the mass ejected directly but the two novae  studied here  are fast novae, which suggests high expansion velocities and lighter mass for the  ejecta \citep{Warner1989,DellaValle2020}.  
Uncertainty in the lithium yield is possible when considering that some Calcium could be in the form of Ca III. 
 \citet{Chugai2020} suggested  the possibility that some over-ionization could be  present in the nova ejecta.
 This would lead to a decrease in the abundance of \ion{Ca}{II}  with the  consequence  to over-estimate the total abundance of $^7$Be. Incidentally, this  would alleviate the  tension between observations and TNR theory \citep{Starrfield2020}.  \citet{Molaro2022MNRAS.509.3258M},  using the photoionization code \texttt{Cloudy} \citep{Ferland2017}, showed that over-ionization in the ejecta of CNe is indeed possible but requires unlikely  low densities and high temperatures of the gas. Moreover, the  detection of neutral species transitions  led them to conclude that the main ionization phase of calcium  is the single-ionized stage. We have  detected the \ion{Mg}{i} 383.8 nm line at the same velocities of the expanding ejecta components of the two novae presented in this work, on Day 16 for ASASSN-190qv and on Day 20 for ASASSN-20ni, respectively. In the spectrum of ASASSN-20ni we also report the presence of \ion{Ca}{I} 422.6 nm line, which suggests a low ionization stage for Calcium. 

\subsection{Historical novae in the LMC}

\begin{figure}
 \includegraphics[width=\columnwidth]{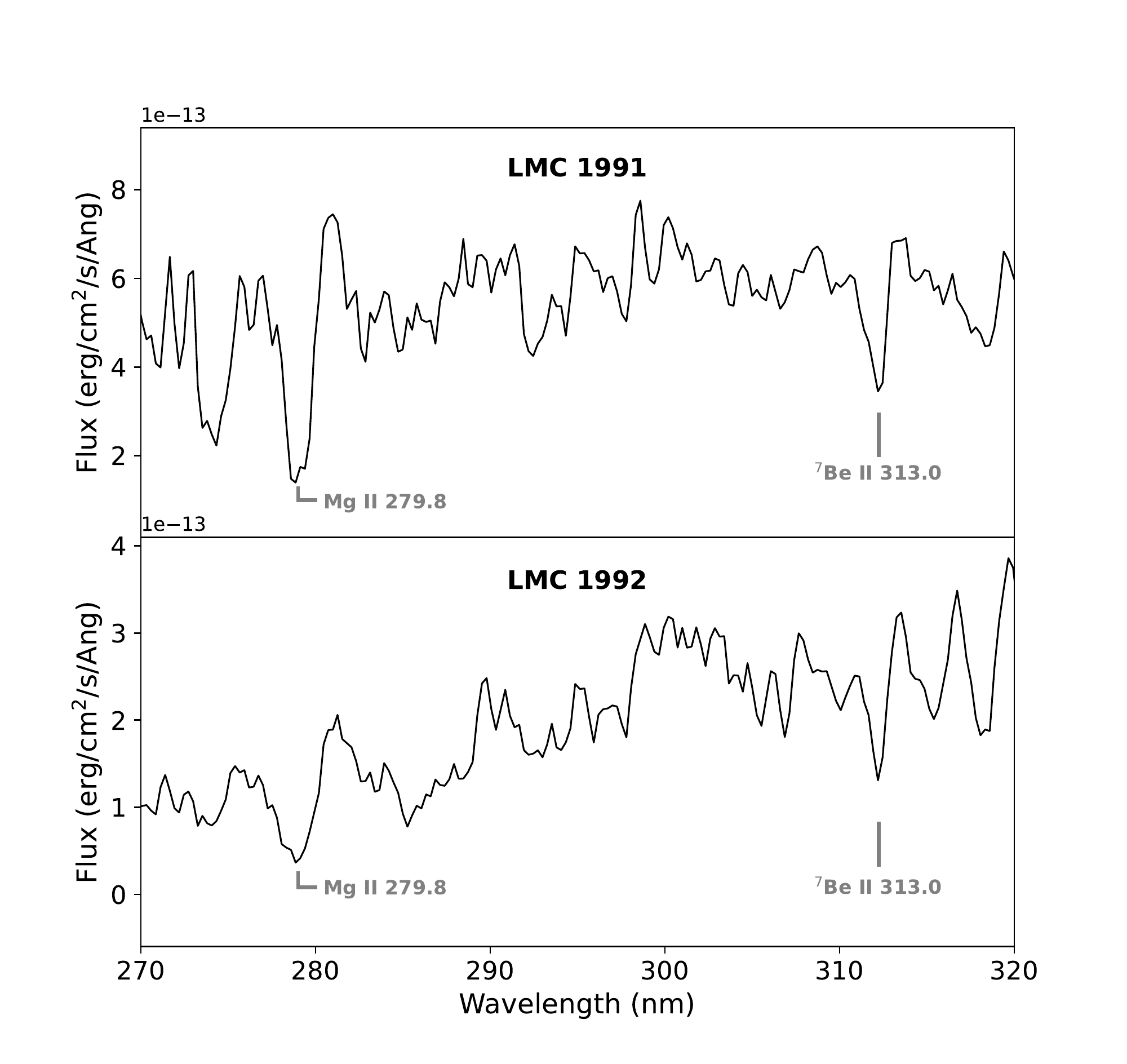}
 \caption{The IUE spectra of LMC 1991 (LWP20210LS, four days after the nova discovery) and LMC 1992 (LWP24303LL, $\sim$one day after the discovery) in the
range 270.0-320.0 nm.  The two strong absorption near 279.0 nm and 312.2 are
identified as Mg II 280.0 and $^7$BeII 313.0 at the expanding blue-shifted velocity of $v_{exp} \sim -850 $ km/s, assuming an expanding velocity for the LMC of $v_{LMC} = 240$ km/s \citep{mcconnachie2012AJ....144....4M}.
Both spectra  are corrected for reddening  using  E(B-V) = 0.15 mag,
see  \citet{cassatella2002A&A...384.1023C}.}
\label{fig:MC}
\end{figure}

Following these detections, we searched in the IUE archive for observations of classical novae exploded in the Magellanic Clouds with IUE, the International Ultraviolet Explorer \citep{Boggess1978}. IUE has already been shown suitable for a $^7$Be search \citep{Selvelli2018}.
The LWP camera of the  IUE satellite  covered the wavelength range
200.0-320.0 nm and  in its low resolution mode, {\sl IUE} ($\Delta\lambda\approx 0.5$\,nm), was  suitable for checking the presence of a wide feature 
near 313.0 nm. 
Nova LMC 1991 and Nova LMC 1992  have been followed by IUE. Representative early spectra for these two novae are shown in Fig. \ref{fig:MC}.
Both spectra exhibit a strong
absorption feature shortward of
$\lambda$313.0 nm  that can be identified  as the blue-shifted
resonance doublet of singly-ionized $^7$\ion{Be}{II}.

We note  that this feature  can be only partially
explained as a blend of common iron curtain absorption  lines of singly
ionized metals. Previous  studies of  galactic novae found only a minor
contribution by singly-ionized metals i.e.  \ion{Cr}{II},
\ion{Fe}{II}, and \ion{Ti}{II} to this feature \citep{Tajitsu2015,Tajitsu2016,Molaro2016,Selvelli2018}, see also Fig. \ref{fig:7}.
A blending  contribution which is expected to be even lower for
Novae in the Large Magellanic Cloud owing   to its  low  metallicity.   \citet{schwarz2001MNRAS.320..103S}, and using  PHOENIX
model atmospheres, found that  the best agreement  between  the
observations and the synthetic spectra of  Nova LMC 1991    requires a
metallicity of Z = 0.1 Z$_{\odot}$. This is a significantly lower metallicity than the
canonical LMC value of  $\sim$ 1/3 solar  and  lends  stronger support to  the
identification  of the $\lambda$313.0 nm feature  as    $^7$\ion{Be}{II}.

Beryllium  and  magnesium  have similar ionization potentials and  a  rough
estimate of the their relative abundances   can be derived from the ratio
of the EWs of their  resonance    absorption doublets. The two lines
show  a
similar
velocity profiles at the same blue-shifted velocity, which indicates that the two features are produced under similar
conditions.
The  observed  absorption EW  of \ion{Mg}{ii} could be
partially reduced  by  the  presence of the emission  component. Moreover, lines of singly-ionized elements, e.g \ion{Cr}{ii}, \ion{Ti}{ii} and \ion{Fe}{ii},  contribute for  10 to 20 percent to 
  the  EW of the $\lambda$313.0 line.
From the data shown in Fig. \ref{fig:MC}   we  measure  an  EW ratio of about 4.2/11.5 = 0.37  in the LMC1991 nova and a 
ratio of 3.6/12.9 = 0.28 in  LMC 1992 nova and  we  adopt  an average  ratio $W(313.0)/W(280.0)$
$\sim$ 0.30.
In the   optically thin regime, this ratio 
provides an estimate of the number of absorbers ($N_{i}$) through  the common
relation of 
$W   \propto   N_i \times f_{ij} \times \lambda^2$.
Since $f_{ij}$(313.0)/$f_{ij}$(280.0)$\sim$0.5   the observed EW ratio provides
$N_i(313.0)/N_i(280.0) =  0.48 $
Since the second ionization potentials of the two ions have similar
values  (i.e., 18.21 eV and 15.04 eV, respectively), similar ionization fractions   are
expected. Thus, the above
derived ratio also  provides  an estimate of the total  $^7$Be/Mg abundance.
Assuming the magnesium abundance being 1/4 of the solar (similar to the value assumed for the SMC, see next Section), namely 9.08 $\times$ 10$^{-6}$  
\citep{Lodders2019},  the $^7$\ion{Be}{ii}~ abundance relative
to hydrogen  is given by
$N(^{7}Be)/N(H) = 9.08 \times10^{-6}\times 0.48  \sim
4.36 \times 10^{-6}$,  which is very close to the yields derived here for 
the novae of the Small Magellanic Cloud.

\subsection{On the Li-enrichment in the SMC}

\citet{Howk2012},  from the detection of the \ion{Li}{I} interstellar line along the line-of-sight  of SK 143,   derived  N($^7$Li)/N(H) = (4.8 $\pm$ 1.8) $\times$ 10$^{-10}$ and an absolute Li abundance of  A(Li) = 2.68 $\pm 0.16$, a value which is   close to the level expected  from the SBBN, when considering the CMB baryon density. \citet{Howk2012} concluded that the value measured in the SMC interstellar medium corresponds to the primordial  with    a negligible stellar post-BBN production.

On the other hand, the nova lithium yield  estimated here suggests a non-negligible role of classical novae in the lithium enrichment of the SMC. The $^7$Be  produced in SMC novae  will enrich  the ISM of the SMC with freshly-produced  $^7$Li. The lithium yield inferred from the analysis of ASASSN-19qv and ASASSN-20ni is $M_{^7 Li} = M_{^7 Be} = (3.7 \pm 0.6) \times 10^{-10}$ M$_{\odot}$. Assuming all SMC novae eject a similar amount of lithium into the ISM of the SMC we can have an approximate estimate of their contribution.
For this purpose we need to know the nova rate in the SMC. This quantity was recently discussed by \citet{DellaValle2020}. These authors found a robust lower limit for the SMC nova rate of 0.7 events per year,  which is consistent with r = 0.9 $\pm$ 0.4 yr${-1}$, measured by \citet{Mroz2016}. In the following we take the latter value of the rate as a constant over the age of the SMC. The oldest globular cluster (GC) in the SMC, NGC 121, is 2-3 Gyr younger than the oldest GCs in the Milky Way \citep{Glatt2008}, and  the average age of stars in the outer regions of the SMC is of $t_{SMC} = 10.6 \pm 0.5$ Gyr  \citep{Dolphin2001}. Considering  a time delay of $\sim 2$ Gyr for the formation of a nova-progenitor WD system,  we  assume  for  the action of nova system in the SMC a time interval $\tau = 8.6 \pm 0.5$ Gyr.  Thus, the total lithium production  from classical novae amounts to:

\begin{equation}
    M_{Li,n} = M_{^7Li}\, \times\, r\, \times\, \tau 
\end{equation}
which gives
\begin{equation}
 M_{Li,n} = (3.7 \pm 0.6) \times 10^{-10} \times 0.9 \times ( 8.6 \pm 0.5) \times 10^9 = (2.8 \pm 1.9) M_{\odot}
\end{equation}
The uncertainties reported in the final estimate of $M_{Li,n}$ correspond to maximum errors.

This value can be compared with Li/H   through  the SMC mass.
An  estimate of the neutral hydrogen gas mass is $M_{HI} = 4.2 \times 10^8$ M$_{\odot}$  by means of the \ion{H}{I} 21cm line observed with the Australian Telescope Compact Array radio telescope \citep{Stanimirovic2004}. This value is  also in agreement with the results from the Parkes \ion{H}{I} survey, $M_{HI} = 4.02 \pm 0.08 \times 10^8$ M$_{\odot}$ \citep{Bruns2005}. The  additional contribution of the molecular hydrogen $H_2$  can been derived from the far-infrared maps provided by the \textit{Spitzer} Survey of the SMC.  \citet{Leroy2007} estimated a total mass of $M_{H_2} = 3.2 \times 10^7$ M$_{\odot}$. Adding  this value twice to the neutral hydrogen mass  we obtain a final value for the hydrogen mass of $M_{H} = (4.8 \pm 0.2) \times 10^8$  M$_{\odot}$. The total mass accounting also for stars was   estimated by \citet{mcconnachie2012AJ....144....4M} in 9.20 $\times 10^8$  M$_{\odot}$.
Therefore,    the atomic fraction Li/H in the SMC  due to novae becomes N($^7$Li)/N(H) = (4.0 $\pm$ 1.5 ) $\times$ 10$^{-10}$, or  A(Li) = 2.64,
which suggests that a non-negligible  fraction of the  lithium observed in the SMC could  be originated from classical nova explosions.

\begin{figure}
 \includegraphics[width=\columnwidth]{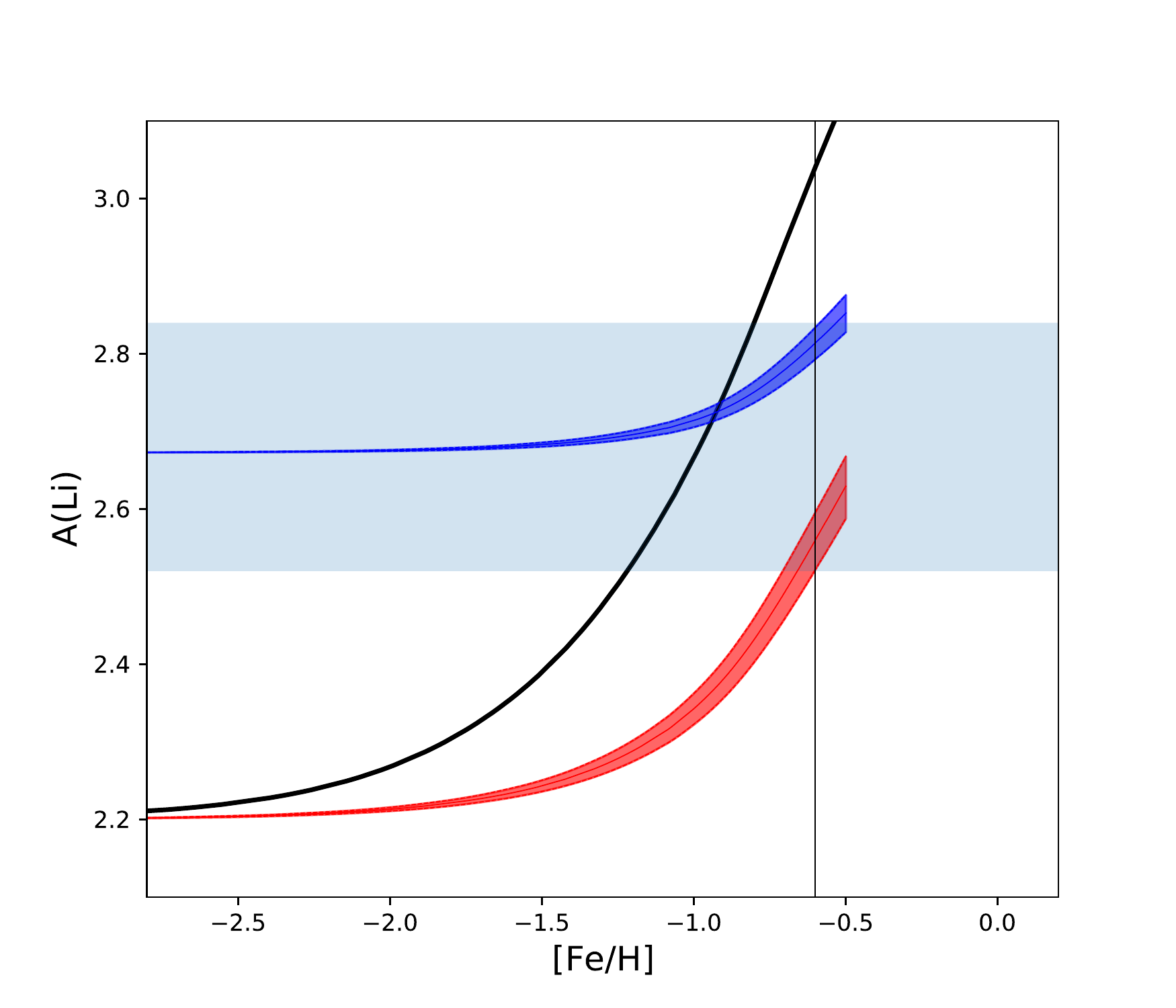}
 \caption{The Li evolution in the SMC assuming the nova yields derived in this paper starting from two different Li initial values: in red the Spite plateau and in blue the CMB+BBN nucleosynthesis. 
 The two curves for each initial lithium abundance and  the derived nova yields with their  error are shown   by  shaded regions. 
 The model include  astration as well as  spallation nucleosynthesis and a modest AGB contribution. The present Li abundance measured in the ISM of SMC  derived by \citet{Howk2012} with  $\pm 1 \sigma$ error is shown with the  shaded horizontal band. Finally, the black curve shows the Li evolution derived when a  chemical evolution of the SMC but with  a Li-yield typical of our Galaxy, namely a factor 4 higher than of the SMC,  is considered \citep{Cescutti2019}. Note how much the expected value for $A(Li)$ would be off with respect to the present Li abundance value in the SMC if we consider MW-like yields.}
 \label{fig:16}
\end{figure}

In Fig. \ref{fig:16} we present  the lithium evolutionary curve for a  detailed chemical model for the SMC. The  model  is similar to those of chemical evolution of  dwarf spheroidal galaxies described in \citet{Lanfranchi2006} as well as for the  Gaia-Enceladus dwarf galaxy described in \citet{Cescutti2020}, the first
model following the evolution of lithium in external galaxies (see also \citealp{Matteucci2021b} for
more satellites of the Milky Way). Compared to  our Galaxy, dwarf galaxies  have lower star formation efficiencies and galactic winds that prevent to reach solar metallicities. In particular, our model for SMC  follows the equations described in \citet{Cescutti2020}
and it has an evolution that lasts 10 Gyr, a Gaussian infall law with a peak at 4 Gyr and sigma of 1 Gyr. The total mass surface density is 50 M$_{\odot}$/pc$^2$ with a star formation efficiency of 0.06 Gyr$^{-1}$.  We assume a Galactic wind that starts after 7 Gyr, with a wind efficiency of 0.8 Gyr$^{-1}$, proportional to the gas still present in the galaxy.  These chemical evolution parameters are fixed to reproduce the metallicity distribution function and the [$\alpha $/Fe] vs [Fe/H] trend observed in giant stars of the SMC by \citet{Mucciarelli2014}. For lithium, we consider the same nucleosynthetic channels  considered in \citet{Cescutti2019}. The Li production  is mainly from novae, with a small contribution from production from AGB \citep{Ventura2010}. For the spallation production, we adopt  the   spallation rates which are empirically derived  from beryllium observations in stars belonging to  the Gaia-Enceladus dwarf galaxy,  which,  however, remains very similar to that of the Galaxy \citep{molaro2020MNRAS.496.2902M}. In Gaia-Enceladus, the relation is 

\begin{equation}
A(Be) =  0.729 (\pm 0.059)\,\times\,  [Fe/H] + 0.856 ( \pm 0.117)
\end{equation}

The scaling of Li/Be $\approx$ 7.6 for spallation processes is adopted \citep{Molaro1997} which provides   at the SMC  metallicity  a value of A(Li) = 1.94  made by spallation processes only. The Li destruction due to astration in the stellar recycling  is also taken into account.
The model is  shown for two different initial Li values, according to the SBBN+Planck  nucleosynthesis of $A(^7Li) = 2.72 \pm 0.06$ \citep{Cyburt2016},  or  the Spite plateau at A(Li) = 2.20$\pm$ 0.05.
The present estimated values are A(Li) = 2.56 $\pm$ 0.04  when starting from the primordial value taken from the halo stars and A(Li)= 2.81 $\pm$ 0.02 for the higher primordial value expected by the SBBN with the baryonic density of the deuterium measurement. These could be  compared with the value  of A(Li) = 2.68 $\pm$ 0.16 measured by \citet{Howk2012} in the interstellar medium. Unfortunately, the present error bar in the ISM determination is so large that it is not possible to discriminate between the two initial values. If this error can be reduced significantly in future observations then the present abundance could discriminate between the two initial values. If the present Li abundance is at the level of the SBBN-Planck Li value this would require a  low primordial value in order to account for  the stellar Li production.  On the contrary,   if the primordial Li  value is  what was indicated by the  SBBN and Planck, then the Li today in the SMC needs to be slightly higher than that due to the contribution of the stellar Li synthesis. At face value the \citet{Howk2012} measurement of A(Li)=2.68 $\pm$ 0.16 favours a low primordial  value.

\section{Conclusions}

After the discovery of ASASSN-19qv in the SMC, we were granted a DDT programme at ESO-VLT with the high-resolution spectrographs UVES and X-shooter, in order to study the spectroscopic evolution of ASASSN-19qv and  to search for the $^7$Be isotope in the early epochs of the nova outburst. One year later, and in the middle of the pandemic, we used granted telescope time to observe the outburst of another nova in the SMC, ASASSN20-ni. The target-of-opportunity nature of our program allowed us to observe the nova very soon after its discovery and  the first high-resolution spectrum with VLT/UVES was obtained  four days later. We summarize the main results here.  

\begin{itemize}
    {\item The $^7$\ion{Be}{ii} resonance transitions were  detected in ASASSN-19qv at two distinct epochs,   and also in  ASASSN-20ni in all the  epochs of the observations.}
    
    {\item We  have analysed the outburst spectra in order to infer the amount of $^7$Be, and therefore $^7$Li, synthesised in TNR of both novae. 
    This  provided the first estimate of the lithium yield by novae in the SMC, namely N($^7$Li)/N(H) = (5.3 $\pm$ 0.2) $\times$ 10$^{-6}$. With a conservative value of $M_{H,ej} =$ 10$^{-5}$ M$_{\odot}$ we have   $M_{^7 Be} = (3.7 \pm 0.6) \times 10^{-10}$ M$_{\odot}$ per nova event.}
    
    {\item We have also studied two historical novae  exploded in the Large Magellanic Cloud   which have been followed by the IUE satellite. The IUE LWP  spectra of LMC 1991 and LMC 1992  exhibit a strong absorption feature  that can be identified  as the  resonance doublet of singly-ionized $^7$\ion{Be}{II}. The low metallicity of the Large Magellanic Cloud make less likely to explain  the feature  as a blend of  absorption  lines of singly
ionized metals.  By using Mg as a reference element we obtain a  $^7$\ion{Be}{ii}~ abundance  of 
$N(^{7}Be)/N(H)  \sim
4.36 \times 10^{-6}$,  which, despite several uncertainties,  is very close to that  derived  for 
the novae of the Small Magellanic Cloud. To note that  Mg  and  $^7$\ion{Be}{ii}~  have   similar  second ionization potentials and of therefore 
the  $^7$\ion{Be}{ii}~ yields obtained by using Mg  are quite insensitive to the presence of overionization effects in the ejecta.  }
    
    {\item When these yields are inserted in a chemical evolution model  suited for  the SMC they result into  a present Li abundance  of the order of A(Li)=2.56 when starting form a low primordial $^7$Li value or A(Li)=2.8 when starting from a high primordial value. The observation of present Li in the interstellar gas of the SMC of \citet{Howk2012} of A(Li)=2.68 $\pm$ 0.16 are  consistent with both values within 1 $\sigma$ error and precludes from any firm conclusion.}
    
     {\item The evidence of $^7$Be in the CN ejecta observed in the Magellanic Clouds suggests that the thermonuclear reactions giving origin to this isotope during the TNR is effective also in environments characterised by a general sub-solar metallicity, as it is indeed the case for the two main Milky Way satellites. All recent CN explosion simulations have always considered a solar abundance value for the material accreted onto the primary WD in a CN binary system \citep{Casanova2018,Jose2020,Starrfield2020}. Therefore, our result implies the necessity of further simulations, characterised by a sub-Solar metallicity for the matter accreting the primary WD, that can support our finding.}
     
     {\item This result implies that CNe are likely the main lithium factories also in systems external to our Galaxy. Further observations will constrain much better the lithium yield in nearby galaxies, such as the SMC and the LMC. This goal will be easily reachable with the possibility to use advanced high-resolution spectrographs observing the near-UV range such as the proposed CUBES at ESO/VLT \citep{CUBES}.  }

\end{itemize}

\section*{Acknowledgements}

We are very grateful to the several dozens of observers worldwide who have contributed with their observations to the AAVSO International Database, a resource that we always consider as a main reference for classical novae evolution light curves. We are also grateful to the ESO director for the DDT programme 2103.D-5044. LI was supported by grants from VILLUM FONDEN (project number 16599 and 25501). AA is supported by a Villum Experiment Grant (project number 36225). This work was partially supported by the European Union (ChETEC-INFRA, project no. 101008324 and ChETEC, CA16117). E.A. acknowledges NSF award AST-1751874, NASA award 11-Fermi 80NSSC18K1746, NASA award 16-Swift 80NSSC21K0173, and a Cottrell fellowship of the Research Corporation. M.H. acknowledges support from 
grant PID2019-108709GB-I00 from MICINN (Spain). We acknowledge Lucas Macri, Jay Strader, and Laura Chomiuk for help with obtaining and reducing the SOAR spectrum. We have also made an extensive use of python scripts, developed specifically for the analysis of this nova, as well as of IRAF tools for a counter-check of all the measurements presented in this work. We recognize the use of the \emph{numpy} \citep{Numpy}, \emph{matplotlib} \citep{Matplotlib}, and the \emph{astropy} \citep{Astropy2013} python packages. Last, but not least, LI is also very grateful to Anna Serena Esposito, given that the first evidence of $^7$Be from a SMC nova was found the same day of our wedding: LI fully recognizes her patience in tolerating the time spent on activating, reducing and analyzing the data within the most beautiful days of our life together. 



\section*{Data Availability}

The data underlying this article will be shared on reasonable request to the corresponding author.



\bibliographystyle{mnras}




\end{document}